\pdfoutput=1
\documentclass[12pt,titlepage]{article}

\font\grande=cmr9.5 scaled \magstep4
\font\medio=cmr9.5 scaled \magstep2
\outer\def\beginsection#1\par{\medbreak\bigskip
      \message{#1}\leftline{\bf#1}\nobreak\medskip
\vskip-\parskip
      \noindent}
\usepackage{graphicx} 
\begin{document}
\bibliographystyle {unsrt}

\titlepage

\begin{flushright}
CERN-TH-2019-090
\end{flushright}

\vspace{1cm}
\begin{center}
{\grande Quantum coherence of relic gravitons}\\
\vspace{0.5cm}
{\grande and Hanbury Brown-Twiss interferometry}\\
\vspace{1cm}
 Massimo Giovannini 
 \footnote{Electronic address: massimo.giovannini@cern.ch} \\
\vspace{1cm}
{{\sl Department of Physics,  CERN, 1211 Geneva 23, Switzerland }}\\
\vspace{0.5cm}
{{\sl INFN, Section of Milan-Bicocca, 20126 Milan, Italy}}
\vspace*{1cm}
\end{center}

\vskip 0.3cm
\centerline{\medio  Abstract}
\vskip 0.1cm
The coherence of the relic gravitons is investigated within a quantum mechanical 
perspective. After introducing the notion and the properties of the generalized Glauber 
correlators valid in the tensor case, the degrees of first- and second-order coherence are 
evaluated both inside and  beyond the effective horizon. The inclusive approach (encompassing 
the polarizations of the gravitons) is contrasted with the exclusive approximation where the
total intensity is calculated either from a single polarization or even from a single mode of 
the field. While the relic gravitons reentering the effective horizon after the end of a quasi-de Sitter 
stage of expansion are first-order coherent, the Hanbury Brown-Twiss correlations 
always exhibit a super-Poissonian statistics with different quantitative features that depend 
on the properties of their initial states and on the average over the tensor polarizations.
\noindent

\vspace{5mm}
\vfill
\newpage
\renewcommand{\theequation}{1.\arabic{equation}}
\setcounter{equation}{0}
\section{Introduction}
\label{sec1}
As gravitational wave astronomy is opening a new observational window, 
the potential implications of the current developments for the stochastic backgrounds of relic gravitons 
are more accurately investigated. In a wide range of scenarios the early 
evolution of the space-time curvature induces a stochastic background of primordial gravitational waves 
with a spectral energy density extending today from frequencies ${\mathcal O}(\mathrm{aHz})$ 
(i.e. $1 \,\mathrm{aHz} = 10^{-18}\, \mathrm{Hz}$) up to frequencies ${\mathcal O}(\mathrm{GHz})$
 (i.e. $1\, \mathrm{GHz} = 10^{9} \mathrm{Hz}$). While the specific features of different models 
will necessarily produce a variety of spectral amplitudes, all the current and planned experiments 
aiming at a direct (or indirect) detection of the relic gravitons are (or will be) sensitive to the average 
intensity of the gravitational field. In the language of the quantum theory of optical coherence the mean
intensity of the field (or the average multiplicity of gravitons) is related to the degree of first-order coherence. 
With the goal of inspiring some of the future endeavours it is interesting to analyze the degrees of quantum coherence of the relic gravitons in a systematic perspective similar to the one already attempted in the case of large-scale curvature inhomogeneities. 

The quantum theory of optical coherence \cite{QO1,QO2,QO3} is customarily formulated in the context of vector fields but it can be generalized to the tensor and scalar cases by appropriately including (or excluding) the relevant polarizations. This is, after all, the logic already followed in quantum optical analyses where the scalar analog of the electromagnetic field is often scrutinized by focussing the attention on a single polarization (see e.g. \cite{QO4}). The  quantum treatment of the problem (not really mandatory prior to the celebrated series of experiments of Hanbury Brown and Twiss \cite{HBT1,HBT2}) stems from the inadequacy of the classical description of the degree of second-order coherence of certain optical fields. According to the same perspective we can argue that the Young two-slit experiment (i.e. first-order correlations) is not a valid criterion to infer the quantum or classical nature of a given radiation field, wether it be a vector field (as in the case of the photons) or a tensor field (as in the case of the gravitons). The interferometry originally developed by Hanbury Brown and Twiss  can be then applied to relic phonons (i.e. quanta corresponding to large-scale curvature perturbations) and relic gravitons as firstly suggested some time ago \cite{HBT3,HBT4}. While various approximations have been attempted, we intend to generalize the Glauber theory to the case of tensor fields. 

The relic gravitons are potentially produced in the early Universe thanks 
to the pumping action of the gravitational field, as suggested in Refs. \cite{GR1,fordp1,fordp2,GR2} even 
prior to the formulation of conventional inflationary models. The quantum theory of parametric 
amplification, originally developed for  photons \cite{PP1,PP2}, has been generalized to the case of fields of different spins and, in particular, to the case of relic gravitons (see e.g. \cite{HBT3,HBT4} and references therein). The gravitons produced from the vacuum or by stimulated emission (i.e. from a specific initial state) typically have opposite comoving momenta and lead to squeezed states \cite{GR3}. Similar 
patterns  arise in different scenarios including the conventional inflationary models  where the spectral energy density has the usual quasi-flat slope \cite{GR4} with a low-frequency break around $100$ aHz \cite{GR5}. Even if a direct detection of relic gravitons is not behind the corner both for technical and conceptual reasons, the degree of quantum coherence of the large-scale correlations could be used to disambiguate their origin, at least in principle \cite{HBT3,GR6,KS}. In the light of these ambitious targets the present investigations are therefore mandatory.
Indeed, the astrophysical events observed by the Ligo/Virgo collaboration  (e.g. the three-detector observations of gravitational waves from black-hole coalescence \cite{LIGO1}, the evidence of gravitational waves from neutron star inspiral \cite{LIGO2}, and the observation of a $50$-solar-mass binary black hole coalescence for a redshift $z =0.2$ \cite{LIGO3}) are qualitatively different from the potential signals coming from 
relic gravitons. Even if the spectra of relic gravitons are rightfully advertised as potential snapshot of the early Universe, their spectral energy density between 
few mHz  (i.e. $1 \,\mathrm{mHz} = 10^{-3}$ Hz) and the kHz is rather minute\footnote{Optimistically ${\mathcal O}(10^{-16.5})$ at least in the case of conventional inflationary scenarios where the absolute normalization of the tensor power spectrum solely depends on the tensor to scalar ratio $r_{T} < 0.07$ \cite{BICPL}. 
The corresponding chirp amplitude is ${\mathcal O}(10^{-29})$ for a comoving 
frequency of ${\mathcal O}(0.1)$ kHz}. The signal may sharply augment when the spectral energy density increases for frequencies larger than the mHz as it happens when the tensor modes of the geometry 
inherit a refractive index \cite{blue1,blue2} or in the presence of stiff phases. In these cases it can happen that
$h_{c} = {\mathcal O}(10^{-25})$ \cite{blue1}, while the chirp amplitude$h_{c}$ corresponding 
to the astronomical signals detected so far by the Ligo/Virgo collaboration is ${\mathcal O}(10^{-21})$ \cite{LIGO1,LIGO2,LIGO3}. Even if the current upper limits on stochastic backgrounds of relic gravitons 
are still far from the final targets \cite{LV3,LV4}, the terrestrial interferometers in their 
advanced version \cite{LV1,LV2} will hopefully probe chirp amplitudes ${\mathcal O}(10^{-25})$ 
corresponding to spectral amplitudes $h_{0}^2 \Omega_{\mathrm{gw}} = {\mathcal O}(10^{-11})$.  
In the foreseeable future the Japanese Kagra (Kamioka Gravitational Wave Detector) \cite{kagra1,kagra2}  (effectively a  prosecution of the Tama-300 experiment \cite{TAMA}) and the Einstein telescope \cite{ET1} should be both operational in the audio band. The GEO-600 detector \cite{GEO1,GEO2} is already 
progressing towards a further reduction of the quantum noise that will probably 
be essential for the third generation of terrestrial wide-band interferometers. 
The  space-borne interferometers, such as (e)Lisa (Laser Interferometer Space Antenna) \cite{LISA}, Bbo (Big Bang Observer) \cite{BBO}, and Decigo (Deci-hertz Interferometer Gravitational Wave Observatory) \cite{DECIGO1,DECIGO2}, should (hopefully) operate between few mHz and the Hz, maybe after 2035. While the sensitivities of these 
instruments are still at the level of targets, we can say that they should probably range 
between $h_{0}^2 \Omega_{\mathrm{gw}} = {\mathcal O}(10^{-12})$ and $h_{0}^2 \Omega_{\mathrm{gw}} = {\mathcal O}(10^{-15})$.

The layout of this investigation is the following. In section \ref{sec2} 
the Glauber correlators are introduced in the tensor case. In section \ref{sec3}  the degree of first-order 
coherence is evaluated both inside and beyond the effective horizon. Section \ref{sec4} is devoted to the 
estimate of the degree of second-order coherence while the role of the initial states will be analyzed in section \ref{sec5}. Section \ref{sec6} contains the concluding discussion. To make the analysis self-contained the most relevant technical results have been relegated to the appendices \ref{APPA} and \ref{APPB}.

\renewcommand{\theequation}{2.\arabic{equation}}
\setcounter{equation}{0}
\section{The quantum coherence of relic gravitons}
\label{sec2}
The canonical theory of optical coherence is formulated in terms of vectors \cite{QO1,QO2,QO3,QO4}
but it is not uncommon to consider the scalar analog of the electromagnetic field by 
discussing a single polarization. The aim of this section is to extend the Glauber approach to the case 
of the divergenceless and traceless tensor fields describing the evolution of the relic gravitons.
\subsection{Glauber correlation functions}
The transverse and traceless fluctuations of the metric are conventionally denoted by $\delta_{t} g_{ij} = - a^2 h_{ij}(x)$ where $x_{i} \equiv (\vec{x}_{i}, \, \tau_{i})$ and $\partial_{i} h^{i}_{\,\,j} = h_{i}^{\,\,i} =0$. 
The background geometry will be taken to be a conformally flat $\overline{g}_{\mu\nu} = a^2(\tau) \eta_{\mu\nu}$ 
where $\eta_{\mu\nu}$ is the Minkowski metric with signature $(+,\, -,\,-,\,-)$ and $a(\tau)$ is 
the scale factor in the conformal time parametrization.  For practical purposes the correlation functions will be defined via the rescaled tensor amplitude $\hat{\mu}_{ij}(x) = \hat{h}_{ij}(x) a(\tau)$ and the hats will denote throughout the quantum field operators as opposed to their classical analog. The operators $\hat{\mu}_{ij}(x)$ consist of a positive and of a negative frequency part, i.e.  $\hat{\mu}_{ij}(x) = \hat{\mu}_{ij}^{(+)}(x) + \hat{\mu}_{ij}^{(-)}(x)$, with 
$\hat{\mu}_{ij}^{(+)}(x)= \hat{\mu}_{ij}^{(-)\,\dagger}(x)$. If  $|\mathrm{vac}\rangle $ is the state that minimizes the tensor Hamiltonian when all the modes are inside the effective horizon (for instance at the onset of inflation) the operator $\hat{\mu}_{ij}^{(+)}(x)$ 
annihilates the vacuum (i.e.  $\hat{\mu}_{ij}^{(+)}(x) |\mathrm{vac} \rangle=0$ and 
$\langle \mathrm{vac} |\, \hat{\mu}_{ij}^{(-)}(x) =0$). In the tensor case the Glauber correlation function is given by: 
\begin{eqnarray}
&& {\mathcal T}^{(n,m)}_{(i_{1}\,\,j_{1}), \,.\,.\,.\,(i_{n}\,\,j_{n}), \, (i_{n+1}\,\,j_{n+1}),\, .\,.\,.\,, (i_{n +m}\,\,j_{n+m}) }(x_{1}, \,.\,.\,.\,x_{n}, \, x_{n+1},\, .\,.\,.\,, x_{n +m}) 
\nonumber\\
&& = \mathrm{Tr}\biggl[ \hat{\rho} \, \hat{\mu}_{i_{1}\,\,j_{1}}^{(-)}(x_{1})\,.\,.\,.\, \hat{\mu}_{i_{n}\,\,j_{n}}^{(-)}(x_{n})
\, \hat{\mu}_{(i_{n+1}\,\,j_{n+1})}^{(+)}(x_{n+1})\,.\,.\,.\,\hat{\mu}_{(i_{n+m}\,\,j_{n +m})}^{(+)}(x_{n+m})\biggr],
\label{corrT1}
\end{eqnarray}
where $\hat{\rho}$ is the density operator representing the (generally mixed) state of the field $\hat{\mu}_{ij}$.  Equation (\ref{corrT1}) generalizes the Glauber correlator (normally written in the case of photons) to the case of gravitons\footnote{Instead of $(n+m)$ {\em vector} indices \cite{QO2}, in Eq. (\ref{corrT1}) we have $(n+m)$ pairs of {\em tensor} indices (i.e. $(i_{1}\, j_{1})\,...\,(i_{n}\, j_{n})\,...\, (i_{n+m}\, j_{n+m})$).}.
In quantum optics an exclusive perspective is often invoked by purposely neglecting one 
of the two polarizations of the photon \cite{QO4}. This choice, motivated by specific empirical 
requirements\footnote{In quantum optics the single-polarization approximation is motivated by  
various experiments dealing with a single polarization (for instance in a cavity); this approach 
is  exclusive since the experiments are typically conceived by only considering a single polarization.
Since in the case of relic gravitons the initial conditions are not observationally accessible,
 it is useful consider also an inclusive approach where the sum over the polarizations is not 
 neglected in the definition of the intensity and the quantum state of relic gravitons is unpolarized.}, amounts to expunging from the Glauber correlators the vector indices. The same logic and notations will be used in the case of the gravitons so that we define the scalar analog of Eq. (\ref{corrT1}):
\begin{eqnarray}
&& {\mathcal S}^{(n,m)}(x_{1}, \,.\,.\,.\,x_{n}, \, x_{n+1},\, .\,.\,.\,, x_{n +m}) 
\nonumber\\
&& = \mathrm{Tr}\biggl[ \hat{\rho} \, \hat{\mu}^{(-)}(x_{1})\,.\,.\,.\, \hat{\mu}^{(-)}(x_{n})
\, \hat{\mu}^{(+)}(x_{n+1})\,.\,.\,.\,\hat{\mu}^{(+)}(x_{n+m})\biggr],
\label{corrT2}
\end{eqnarray}
with the proviso that now Eq. (\ref{corrT2}) holds in the case of a single tensor polarization. 
It is finally not uncommon to treat the Mach-Zehnder and Hanbury Brown-Twiss interferometry in terms of  a {\em single mode} of the field \cite{loudon}. The single-mode experiments use plane parallel light beams whose transverse intensity profiles are not important for the measured quantities. In these situations is often sufficient to consider the light beams as exciting a single mode of the field. 
This viewpoint is even more exclusive than the one described by  Eq. (\ref{corrT2}) where all the modes of the field are taken into account. In the explicit estimates of the following sections we shall consider, in this order,  the inclusive description of Eq. (\ref{corrT1}), the exclusive approach of the single-polarization approximation (i.e. Eq. (\ref{corrT2}))  and 
finally the single-mode approximation.

\subsection{Physical interpretations of the Glauber correlators}
Equations (\ref{corrT1}) and (\ref{corrT2})  arise when considering the $n$-fold delayed  coincidence measurement of the tensor field at the space-time points $(x_{1}, \,.\,.\,.\,x_{n}, \, x_{n})$. Let us focus, in particular, on the operator appearing inside the trace of Eq. (\ref{corrT1}), namely
\begin{equation}
\hat{{\mathcal Q}}_{(i_{1}\,\,j_{1}), \,.\,.\,.\,(i_{n}\,\,j_{n})}(x_{1}, \,.\,.\,.\,x_{n}) = \hat{\mu}^{(-)}_{i_{1}\,\,j_{1}}(x_{1})\,.\,.\,.\, \hat{\mu}_{i_{n}\,\,j_{n}}^{(-)}(x_{n})
\, \hat{\mu}_{i_{1}\,\,j_{1}}^{(+)}(x_{1})\,.\,.\,.\, \hat{\mu}_{i_{n}\,\,j_{n}}^{(+)}(x_{n}),
\label{corrT3}
\end{equation}
and let us define $|\{\,a\}\rangle$ as the state of the field  after the measurement 
and  $|\{\,b\}\rangle$ the state of the field  before the measurement. The matrix element corresponding to the 
absorption of gravitons at different times and at different locations of the hypothetical detectors 
can then be expressed as $\langle \{a\} \, | \hat{\mu}_{i_{1}\,\,j_{1}}^{(+)}(x_{1})\,.\,.\,.\, \hat{\mu}_{i_{n}\,\,j_{n}}^{(+)}(x_{n})|\, \{b\}\rangle$. 
To obtain the rate at which the absorptions occur we must sum over the final states, i.e. 
\begin{eqnarray}
&& \sum_{\{a\}} \biggl|\langle \{ a\}\, | \hat{\mu}_{i_{1}\,\,j_{1}}^{(+)}(x_{1})\,.\,.\,.\, \hat{\mu}_{i_{n}\,\,j_{n}}^{(+)}(x_{n})| \{ b\} \rangle \biggr|^2 =
 \nonumber\\
&& \equiv \sum_{\{a\}} \langle \{ b\}| \hat{\mu}_{i_{1}\,\,j_{1}}^{(-)}(x_{1})\,.\,.\,.\, \hat{\mu}_{i_{n}\,\,j_{n}}^{(-)}(x_{n})| \{a\}\rangle \langle \{ a \}| 
 \hat{\mu}_{i_{1}\,\,j_{1}}^{(+)}(x_{1})\,.\,.\,.\, \hat{\mu}_{i_{n}\,\,j_{n}}^{(+)}(x_{n}) |\{ b \}\rangle,
\label{corrT4}
\end{eqnarray}
that coincides, thanks to the completeness relation, with the expectation value $\langle \{\, b\}| [\,.\,.\,.] |\{ b\} \rangle$ where the ellipses stand for the operator given in Eq. (\ref{corrT3}). When the operator of Eq. (\ref{corrT3}) is averaged over the ensemble of the initial states of the system it becomes practically identical to Eq. (\ref{corrT1}) for $x_{n + r} = x_{r}$ and $(i_{n+r}\, j_{n +r}) = (i_{r}\, j_{r})$  with $r= 1, 2, \,.\,.\,., n$ and $n=m$. The Glauber correlation function in the tensor case will therefore correspond to:
\begin{eqnarray}
&&{\mathcal T}^{(n)}_{(i_{1}\,\,j_{1}), \,.\,.\,.\,(i_{n}\,\,j_{n})\,.\,.\,.\,(i_{2n}\,\,j_{2n})}(x_{1}, \,.\,.\,.\,x_{n}, \, x_{n+1},\, .\,.\,.\,, x_{2n}) 
\nonumber\\
&&= \mathrm{Tr}\biggl[ \hat{\rho} \, \hat{\mu}^{(-)}_{(i_{1}\,\,j_{1})}(x_{1})\,.\,.\,.\, \hat{\mu}^{(-)}_{(i_{n}\,\,j_{n})}(x_{n})
\, \hat{\mu}_{(i_{n+1}\,\,j_{n+1})}^{(+)}(x_{n+1})\,.\,.\,.\, \hat{\mu}_{(i_{2n}\,\,j_{2n})}^{(+)}(x_{2n})\biggr].
\label{corrT5}
\end{eqnarray}
The analog of Eq. (\ref{corrT5}) in the single-polarization approximation follows instead from Eq. (\ref{corrT2})
and it is given by:
\begin{equation}
{\mathcal S}^{(n)}(x_{1}, \,.\,.\,.\,x_{n}, \, x_{n+1},\, .\,.\,.\,, x_{2n}) = \mathrm{Tr}\biggl[ \hat{\rho} \, \hat{\mu}^{(-)}(x_{1})\,.\,.\,.\, \hat{\mu}^{(-)}(x_{n})
\, \hat{\mu}(x_{n+1})\,.\,.\,.\, \hat{\mu}(x_{2n})\biggr].
\label{corrT6}
\end{equation}
Having extended the Glauber correlators to the tensor case, it is now useful to introduce the corresponding degrees of quantum coherence.

\subsection{Degrees of coherence for tensor fields}
The first-order Glauber correlation function follows 
from Eq. (\ref{corrT5}) for $n=1$ and it is:
\begin{equation}
{\mathcal T}^{(1)}_{(i_{1}\,\,j_{1}),\,(i_{2}\,\,j_{2})}(x_{1},\, x_{2}) = \langle  \hat{\mu}^{(-)}_{i_{1}\,\,j_{1}}(x_{1}) \, \hat{\mu}^{(+)}_{i_{2}\,\,j_{2}}(x_{2}) \rangle.
\label{corrT7a}
\end{equation}
From Eq. (\ref{corrT6}) we can similarly obtain the analog of Eq. (\ref{corrT7a}) in the 
single-polarization approximation:
\begin{equation}
{\mathcal S}^{(1)}(x_{1},\, x_{2}) = \langle  \hat{\mu}^{(-)}(x_{1}) \, \hat{\mu}^{(+)}(x_{2}) \rangle.
\label{corrT7b}
\end{equation} 
When $(i_{1}\,\,j_{1}) = (i_{2}\,\,j_{2})$ Eq. (\ref{corrT7a}) describes the intensity averaged over the tensor 
polarizations and shall be denoted by 
\begin{equation}
{\mathcal T}^{(1)}(x_{1},\, x_{2}) =  \langle  \hat{\mu}^{(-)}_{i\,j}(x_{1}) \, \hat{\mu}^{(+)}_{i\,j}(x_{2}) \rangle.
\label{corrT7c}
\end{equation}
Equation (\ref{corrT5}) in the case $n=2$ determines 
the second-order correlation function which is relevant when discussing the Hanbury Brown-Twiss 
interferometry in the tensor case:
\begin{equation}
{\mathcal T}^{(2)}_{(i_{1}\,\,j_{1}), \,(i_{2}\,\,j_{2})\,(i_{3}\,\,j_{3}),\,(i_{4}\,\,j_{4}) }(x_{1}, x_{2}, x_{3}, x_{4}) = 
 \langle  \hat{\mu}^{(-)}_{i_{1}\,\,j_{1}}(x_{1}) \, \hat{\mu}^{(-)}_{i_{2}\,\,j_{2}}(x_{2})\hat{\mu}^{(+)}_{i_{3}\,\,j_{3}}(x_{3}) \hat{\mu}^{(+)}_{i_{4}\,\,j_{4}}(x_{4}) \rangle.
\label{corrT8a}
\end{equation} 
Similarly from Eq. (\ref{corrT6}) we can obtain the analog of Eq. (\ref{corrT8a}) 
in the single-polarization approximation:
\begin{equation}
{\mathcal S}^{(2)}(x_{1}, x_{2}, x_{3}, x_{4}) = \langle  \hat{\mu}^{(-)}(x_{1}) \, \hat{\mu}^{(-)}(x_{2})\hat{\mu}^{(+)}(x_{3}) \hat{\mu}^{(+)}(x_{4}) \rangle.
\label{corrT8b}
\end{equation}
Equations (\ref{corrT8a}) and (\ref{corrT8b}) can describe the correlations of the intensities at two 
separate space-time points. This choice corresponds to the interferometric strategy pioneered by Hanbury Brown and Twiss (HBT) \cite{HBT1,HBT2} as opposed to the standard Young-type experiments where only amplitudes (rather than intensities) are allowed to interfere.
The applications of the HBT  ideas range from stellar astronomy \cite{HBT1,HBT2}  to subatomic 
physics \cite{revs}. The interference of the intensities has been used to determine the hadron 
fireball dimensions \cite{cocconi1,cocconi2} corresponding to the linear size of the 
interaction region in proton-proton collisions. To disambiguate the possible origin of large-scale  curvature perturbations \cite{HBT3,HBT4} and of relic gravitons, probably the only hope is the analysis of the degree of second-order coherence, as we shall argue. Since the intensity must be Hermitian \cite{loudon} the standard HBT correlators 
follow from Eqs.  (\ref{corrT8a}) and (\ref{corrT8b}) by requiring 
\begin{equation}
x_{2} = x_{3}, \qquad (i_{2}\,\,j_{2}) =  (i_{3}\,\,j_{3}), \qquad x_{4}=  x_{1}, \qquad (i_{1}\,\,j_{1}) =  (i_{4}\,\,j_{4}).
\label{corrT9}
\end{equation}
Thus with the identifications (\ref{corrT9}) the intensity correlators will be given by:
\begin{eqnarray}
&& {\mathcal T}^{(2)}(x_{1}, x_{2}) = 
 \langle  \hat{\mu}^{(-)}_{i\,\,j}(x_{1}) \, \hat{\mu}^{(-)}_{k\,\,\ell}(x_{2})\hat{\mu}^{(+)}_{k\,\ell}(x_{2}) \hat{\mu}^{(+)}_{i\,j}(x_{1}) \rangle, 
\label{corrT9a}\\
&& {\mathcal S}^{(2)}(x_{1}, x_{2}) = \langle  \hat{\mu}^{(-)}(x_{1}) \, \hat{\mu}^{(-)}(x_{2})\hat{\mu}^{(+)}(x_{2}) \hat{\mu}^{(+)}(x_{1}) \rangle.
\label{corrT9b}
\end{eqnarray}
Note that Eq. (\ref{corrT9a}) does depend on the polarizations through the sum over the tensor indices while 
the sum does not appear in Eq. (\ref{corrT9b}). From the results of Eqs. (\ref{corrT7a})--(\ref{corrT7c}) 
and of Eqs. (\ref{corrT9})--(\ref{corrT9b}) the corresponding degrees of quantum coherence 
can be easily obtained. More specifically the degrees of first-order coherence are:
\begin{eqnarray}
g^{(1)}(x_{1},\, x_{2}) = \frac{{\mathcal T}^{(1)}(x_1,\, x_{2})}{\sqrt{{\mathcal T}^{(1)}(x_1)} \,\, \sqrt{{\mathcal T}^{(1)}(x_2)}}, \qquad 
\overline{g}^{(1)}(x_{1},\, x_{2}) = \frac{{\mathcal S}^{(1)}(x_1,\, x_{2})}{\sqrt{{\mathcal S}^{(1)}(x_1)} \,\, \sqrt{{\mathcal S}^{(1)}(x_2)}}.
\label{corrT10a}
\end{eqnarray}
When the degree of first-order coherence has an overline it means 
that it is evaluated in the single-polarization approximation. The single-mode approximation 
for the degrees of coherence will be distinguished by a subscript (i.e. $g^{(1)}_{s}$).
For the sake of conciseness in Eq. (\ref{corrT10a}) the following notations have been used:
\begin{eqnarray}
{\mathcal T}^{(1)}(x_1) &=& {\mathcal T}^{(1)}(x_1,\, x_{1}), \qquad 
{\mathcal T}^{(1)}(x_2)= {\mathcal T}^{(1)}(x_2,\, x_{2}),
\nonumber\\
{\mathcal S}^{(1)}(x_1) &=& {\mathcal S}^{(1)}(x_1, x_{1}), \qquad 
{\mathcal S}^{(1)}(x_2)= {\mathcal S}^{(1)}(x_2, x_{2}),
\label{corrT10c}
\end{eqnarray}
and the same notations spelled out in Eq. (\ref{corrT10c}) will be used throughout.
Finally the  degrees of second-order coherence for the relic gravitons follow from Eqs. (\ref{corrT9a}) and (\ref{corrT9b}) and they are given by:
\begin{eqnarray}
g^{(2)}(x_{1},\, x_{2}) = \frac{{\mathcal T}^{(2)}(x_1,\, x_{2})}{{\mathcal T}^{(1)}(x_1) \,\, {\mathcal T}^{(1)}(x_2)},
\qquad 
\overline{g}^{(2)}(x_{1},\, x_{2}) = \frac{{\mathcal S}^{(2)}(x_1,\, x_{2})}{{\mathcal S}^{(1)}(x_1) \,\, {\mathcal S}^{(1)}(x_2)}.
\label{corrT11a}
\end{eqnarray}
As in the case of the degree of first-order coherence the overline refers to the 
case of a single polarization; as usual $g_{s}^{(2)}$ 
 will denote the degree of second-order coherence in the single-mode approximation.
\renewcommand{\theequation}{3.\arabic{equation}}
\setcounter{equation}{0}
\section{First-order coherence of relic gravitons}
\label{sec3}
The field operators describing the positive and negative frequency parts can be expressed as:
\begin{eqnarray}
\hat{\mu}^{(-)}(\vec{x}, \tau) &=& \frac{\sqrt{2} \ell_{P}}{(2\pi)^{3/2}} \sum_{\alpha} \int \frac{d^{3} k}{\sqrt{2 k}} e^{(\alpha)}_{ij} \, \hat{a}_{- \vec{k}\, \alpha}^{\dagger}(\tau) \, e^{- i \vec{k}\cdot\vec{x}},
\label{degA1}\\
\hat{\mu}^{(+)}(\vec{x}, \tau) &=& \frac{\sqrt{2} \ell_{P}}{(2\pi)^{3/2}} \sum_{\alpha}
\int \frac{d^{3} k}{\sqrt{2 k}} e^{(\alpha)}_{ij} \, 
\hat{a}_{\vec{k}\, \alpha}(\tau) e^{- i \vec{k}\cdot\vec{x}},
\label{degA2}
\end{eqnarray}
where $\ell_{P} = \sqrt{8 \pi G}$ and $e^{(\alpha)}_{ij}$ (with $\alpha = \otimes, \,\oplus$) denotes the polarization tensor
of the graviton. The creation and annihilation operators obey $[\hat{a}_{\vec{k}\,\alpha},\, \hat{a}^{\dagger}_{\vec{p}\,\beta}] = \delta^{(3)}(\vec{k} - \vec{p}) \delta_{\alpha\beta}$ and their evolution  follows from the quantum Hamiltonian discussed in appendix 
\ref{APPA} and here reported in the absence of coherent component\footnote{The coherent component will be 
seperately analyzed in section \ref{sec5}. To avoid digressions, some technical aspects involving the notational conventions have been relegated to the appendix \ref{APPA}.}: 
\begin{equation}
\hat{H}^{(t)} = \frac{1}{2} \int d^{3} p \sum_{\alpha} \biggl\{ p \biggl[ \hat{a}^{\dagger}_{\vec{p}\,\,\alpha} \hat{a}_{\vec{p}\,\,\alpha} + \hat{a}_{- \vec{p}\,\,\alpha} \hat{a}^{\dagger}_{-\vec{p}\,\,\alpha} \biggr] 
+\lambda \hat{a}^{\dagger}_{-\vec{p}\,\,\alpha} \hat{a}_{\vec{p}\,\,\alpha}^{\dagger} 
+ \lambda^{*} \hat{a}_{\vec{p}\,\,\alpha} \hat{a}_{-\vec{p}\,\,\alpha} \biggr\}, 
\label{degA3}
\end{equation}
where $\lambda = i {\mathcal H} = i a^{\prime}/a$ and, as already mentioned, the prime will denote throughout the discussion a derivation with respect to the conformal time coordinate. Equation (\ref{degA3}) includes the sum over the two polarizations of the gravitons and is the continuous-mode 
generalization of the quantum mechanical Hamiltonian of Mollow and Glauber \cite{PP1} (see also \cite{sq1}). In the analysis of the evolution of the scalar and tensor modes of the geometry 
the quantum optical analogy has been firstly pointed out in Ref. \cite{GR3,sq2}. 

\subsection{General expressions for the degree of first-order coherence}
The evolution of  $\hat{a}_{\vec{k},\,\alpha}(\tau)$ and  $\hat{a}_{-\vec{k},\,\alpha}^{\dagger}(\tau)$ 
is determined by the Hamiltonian of Eq. (\ref{degA3}) via the corresponding equations in the Heisenberg 
representation (see   Eq. (\ref{AA9})). Their solution is then expressed in terms of the values of the 
creation and annihilation operators at the reference time $\tau_{i}$:
 \begin{eqnarray}
\hat{a}_{\vec{p},\,\alpha}(\tau) &=& u_{p,\,\alpha}(\tau,\tau_{i}) \hat{b}_{\vec{p},\,\alpha}(\tau_{i}) -  
v_{p,\,\alpha}(\tau,\tau_{i}) \hat{b}_{-\vec{p},\,\alpha}(\tau_{i}),  
\label{degA4}\\
\hat{a}_{-\vec{p},\,\alpha}^{\dagger}(\tau) &=& u_{p,\,\alpha}^{*}(\tau,\tau_{i}) \hat{b}_{-\vec{p},\,\alpha}^{\dagger}(\tau_{i})  -  v_{p,\,\alpha}^{*}(\tau,\tau_{i}) \hat{b}_{\vec{p},\,\alpha}^{*}(\tau_{i}),
\label{degA5}
\end{eqnarray}
where the subscript $\vec{p}$ denotes the comoving three-momentum while the subscript $\alpha$ refers to the polarization. All the wavelengths that are today of the order of (or smaller 
than) the Hubble radius were presumably inside the effective horizon at $\tau_{i}$  (i.e. $ k \tau_{i} \gg 1$)
as it happens, for instance, in the case of conventional inflationary models.
The two complex functions $u_{p,\,\alpha}(\tau,\tau_{i})$ and $v_{p,\,\alpha}(\tau,\tau_{i})$ appearing in Eqs. (\ref{degA4}) and (\ref{degA5}) are then solely determined by the specific dynamical evolution of the pump field $\lambda$ 
 appearing in Eq. (\ref{degA3}) but they are also subjected to the condition $|v_{p,\,\alpha}(\tau,\tau_{i})|^2 - |u_{p,\,\alpha}(\tau,\tau_{i})|^2 =1$ since  the commutation relations between the two sets of creation and annihilation 
 operators must be preserved. For each of the two tensor polarizations,  $u_{p,\,\alpha}(\tau,\tau_{i})$ and $v_{p,\,\alpha}(\tau,\tau_{i})$ depend upon one amplitude and two phases and in spite of their specific unitary evolution they can always be parametrized as: 
\begin{equation}
u_{k,\,\alpha}(\tau,\tau_{i}) = e^{- i\, \delta_{k,\,\alpha}} \cosh{r_{k,\,\alpha}}, 
\qquad v_{k,\,\alpha}(\tau,\tau_{i}) =  e^{i (\theta_{k,\,\alpha} + \delta_{k,\,\alpha})} \sinh{r_{k,\,\alpha}},
\label{degA9}
\end{equation}
where $\delta_{k,\,\alpha}$, $\theta_{k,\,\alpha}$ and $r_{k\,\alpha}$ are all real and depend on $\tau$ and $\tau_{i}$.  The canonical transformation of Eqs. (\ref{degA4}) and (\ref{degA5}) is generated by the squeezing operator ${\mathcal S}(z)$ and by the rotation operator ${\mathcal R}(\delta)$:
\begin{equation}
{\mathcal S}(z) = e^{\sigma(z)/2}, \qquad {\mathcal R}(\delta)= e^{- i n(\delta)/2},
\label{degA9a}
\end{equation}
where $\sigma(z)$ and $n(\delta)$ involve both an integral over the modes and a sum over the two tensor 
polarizations\footnote{These two operators are typically expressed for a discrete set of 
modes but, in the present context, their continuous-mode generalization must be considered as it happens in the derivation of the ground state wavefunction of an interacting Bose gas at zero temperature \cite{fetter,solomon}; the same approach has been used to describe the superfluid ground state \cite{valatin1,valatin2}}:
\begin{eqnarray}
\sigma(z) &=& \sum_{\lambda} \int d^{3} k \,\, \biggl[ z_{k\,\,\lambda}^{*} \, \hat{b}_{\vec{k},\,\lambda} 
\hat{b}_{-\vec{k},\,\lambda} - z_{k,\,\lambda} \, \hat{b}^{\dagger}_{-\vec{k},\,\lambda} 
\hat{b}^{\dagger}_{\vec{k},\,\lambda} \biggr],
\label{BB2}\\
n(\delta) &=& \sum_{\lambda}  \int d^{3}k\, \delta_{k,\,\lambda}\,\biggl[ \hat{b}_{\vec{k},\,\lambda}^{\dagger} 
 \hat{b}_{\vec{k},\,\lambda} + \hat{b}_{-\vec{k},\,\lambda}  \hat{b}_{-\vec{k},\,\lambda}^{\dagger} \biggr].
 \label{BB3}
 \end{eqnarray}
 Note that $\delta_{k,\lambda}$ is real while $z_{k,\, \lambda} = r_{k,\,\lambda} e^{i \theta_{k,\,\lambda}}$;
the parametrization of Eq. (\ref{degA9}) follows from Eq. (\ref{degA4}) and (\ref{degA5}) 
by appreciating that\footnote{Note, incidentally, that $\delta_{k,\,\alpha}$ denotes the phase $\delta$ with modulus of three-momentum 
$k$ and polarization $\alpha$; this quantity has noting to do with the Kroeneker $\delta_{\alpha\beta}$ where 
$\alpha$ and $\beta$ are instead two generic tensor polarizations. With this specification 
no confusion is possible.}:
 \begin{eqnarray}
{\mathcal R}^{\dagger}(\delta) \, {\mathcal S}^{\dagger}(z) \, \hat{b}_{\vec{k},\,\alpha} 
{\mathcal S}(z) \, {\mathcal R}(\delta) 
= e^{- i\, \delta_{k,\,\alpha}} \cosh{r_{k\,\,\alpha}} \hat{b}_{\vec{k},\,\alpha} -  e^{i (\theta_{k,\,\alpha} + \delta_{k,\,\alpha})} \sinh{r_{k,\,\alpha}} \hat{b}^{\dagger}_{-\vec{k},\,\alpha},
\label{degA8a}\\
{\mathcal R}^{\dagger}(\delta) \, {\mathcal S}^{\dagger}(z) \, \hat{b}_{-\vec{k},\,\alpha}^{\dagger} 
{\mathcal S}(z) \, {\mathcal R}(\delta) 
= e^{i\, \delta_{k,\,\alpha}} \cosh{r_{k\,\,\alpha}} \hat{b}^{\dagger}_{-\vec{k},\,\alpha} -  e^{- i (\theta_{k,\,\alpha} + \delta_{k,\,\alpha})} \sinh{r_{k,\,\alpha}} \hat{b}_{\vec{k},\,\alpha}.
\label{degA8b}
\end{eqnarray}
Recalling the specific form of Eqs. (\ref{BB2}) and (\ref{BB3}), the squeezed states of the field will then be denoted as $|\{ z \, \delta \}\rangle = {\mathcal S}(z) \, {\mathcal R}(\delta)\, |\mathrm{vac}\rangle$. From the Hamiltonian (\ref{degA3}) the evolution equations for $r_{k,\,\lambda}$, $\theta_{k,\lambda}$ and $\delta_{k,\, \lambda}$ can be easily derived and are reported in Eqs. (\ref{BB8a}) and (\ref{BB8b}).
Inserting Eqs. (\ref{degA1})--(\ref{degA2}) into Eq. (\ref{corrT7a})  and using Eqs. (\ref{degA4}) and (\ref{degA5}) we obtain the explicit 
form of the first-order Glauber correlation function given in Eq. (\ref{corrT7a}):
\begin{eqnarray}
{\mathcal T}^{(1)}_{i j k \ell}(x_{1}, \, x_{2}) &=& \langle \hat{\mu}_{ij}^{(-)}(x_{1}) \, \hat{\mu}_{ij}^{(+)}(x_{2}) \rangle
\nonumber\\
&=& \frac{1}{(2\pi)^3} \int \frac{d^{3} k}{\sqrt{2 k}} \int \frac{d^{3} p}{\sqrt{2 p}} \, \sum_{\alpha} \sum_{\beta} \, e^{(\alpha)}_{ij}\, e^{(\beta)}_{k\ell} \, e^{ - i (\vec{k}\cdot\vec{x}_{1} + \vec{p}\cdot\vec{x}_2)}
\nonumber\\
&\times& 
v_{k,\,\alpha}^{*}(\tau_{1}, \tau_{i}) v_{k,\,\beta}^{*}(\tau_{2}, \tau_{i}) \langle \hat{b}_{\vec{k},\,\alpha}(\tau_{i}) 
\hat{b}_{-\vec{p},\,\beta}^{\dagger}(\tau_{i}) \rangle.
\label{degA10}
\end{eqnarray}
If the operator  $\hat{b}_{\vec{k},\,\alpha}(\tau_{i})$ annihilates the initial state at $\tau_{i}$ 
the expectation value appearing in the last line of Eq. (\ref{degA10}) corresponds to $\delta^{(3)}(\vec{k} + \vec{p}) \, \delta_{\alpha\beta}$. However, to account for the possible presence of a finite number 
of gravitons at $\tau_{i}$, the expectation value shall be modified as
$\langle \hat{b}_{\vec{k},\,\alpha}(\tau_{i}) 
\hat{b}_{-\vec{p},\,\beta}^{\dagger}(\tau_{i}) \rangle = [ \overline{n}_{k}(\tau_{i}) + 1] \delta^{(3)}(\vec{k} + \vec{p}) \, \delta_{\alpha\beta}$ where $\overline{n}_{k}(\tau_{i})$ denotes the average multiplicity of the initial state. With these specifications, Eq. (\ref{degA10}) becomes: 
\begin{equation}
{\mathcal T}^{(1)}_{i j k \ell}(x_{1}, \, x_{2}) = \frac{1}{(2\pi)^3} \int \frac{d^{3} k}{2 k} \sum_{\alpha} e^{(\alpha)}_{ij} e^{(\alpha)}_{k\ell} 
v_{k,\,\alpha}^{*}(\tau_{1}, \tau_{i}) v_{k,\,\alpha}^{*}(\tau_{2}, \tau_{i}) [\overline{n}_{k}(\tau_{i}) +1 ] \, e^{- i \vec{k}\cdot \vec{r}},
\label{degA11}
\end{equation}
where $\vec{r}= (\vec{x}_{1} - \vec{x}_{2})$. In the standard situation  the relic graviton background is not polarized so that the $u_{k,\alpha}$ and $v_{k,\alpha}$ are the same for each of the two polarizations:
 \begin{equation}
 v_{k,\otimes}(\tau,\tau_{i}) = v_{k,\oplus}(\tau, \tau_{i}) = v_{k}(\tau, \tau_{i}), \qquad 
 u_{k,\otimes}(\tau,\tau_{i}) = u_{k,\oplus}(\tau, \tau_{i}) = u_{k}(\tau, \tau_{i}). 
 \label{degA11a}
 \end{equation}
Inserting the condition (\ref{degA11a}) into Eq. (\ref{degA11}) we have that the first-order 
correlator becomes: 
 \begin{eqnarray} 
 {\mathcal T}^{(1)}_{i j k \ell}(\vec{x}_{1},\, \vec{x}_{2}; \tau_{1},\, \tau_{2}) &=& \frac{2}{(2\pi)^3} \int \frac{d^{3} k}{k} {\mathcal A}_{i j k \ell}(\hat{k}) 
 v_{k}^{*}(\tau_{1}, \tau_{i}) v_{k}^{*}(\tau_{2}, \tau_{i}) [\overline{n}_{k}(\tau_{i}) +1 ] \, e^{- i \vec{k}\cdot \vec{r}},
\label{degA12}\\ 
{\mathcal A}_{i j k \ell}(\hat{k}) &=& \frac{1}{4} \biggl[p_{i k}(\hat{k}) \, p_{j \ell}(\hat{k}) + p_{i \ell}(\hat{k})  p_{j k}(\hat{k}) - p_{i j}(\hat{k}) p_{k \ell}(\hat{k})\biggr],
\label{degA13}
\end{eqnarray}
where $p_{i j}(\hat{k}) = [\delta_{i j} - \hat{k}_{i} \hat{k}_{j}]$. The degree of first-order coherence is determined by Eq. (\ref{degA13}) with $k= i$ and $\ell = j$:
\begin{equation}
{\mathcal T}^{(1)}(\vec{x}_{1},\, \vec{x}_{2}; \tau_{1},\, \tau_{2})= {\mathcal T}^{(1)}_{i j i j}(\vec{r}; \tau_{1},\, \tau_{2}) 
= \frac{1}{\pi^2}\int k d k \,j_{0}(k r) \,  v_{k}^{*}(\tau_{1}, \tau_{i}) v_{k}(\tau_{2}, \tau_{i}) [\overline{n}_{k}(\tau_{i}) +1 ],
\label{degA14}
\end{equation}
where $r= |\vec{x}_{1} - \vec{x}_{2}|$ and $j_{0}(k r) = \sin{k r}/(k r)$ is the zeroth-order spherical Bessel function \cite{abr1,abr2}. In the single-polarization approximation the analog of Eq. (\ref{degA14}) reads
\begin{equation}
{\mathcal S}^{(1)}(\vec{x}_{1},\, \vec{x}_{2}; \tau_{1},\, \tau_{2}) = \frac{1}{4} {\mathcal T}^{(1)}(\vec{x}_{1},\, \vec{x}_{2}; \tau_{1},\, \tau_{2}).
\label{degA15}
\end{equation}
In Eq. (\ref{degA15})  the factor $4$ comes from the sum over the polarizations that is counted in ${\mathcal T}^{(1)}(\vec{r}; \tau_{1},\, \tau_{2})$ but not in ${\mathcal S}^{(1)}(\vec{r}; \tau_{1},\, \tau_{2}) $. As a consequence 
 the degree of first-order coherence is
\begin{eqnarray}
&& g^{(1)}(r; \tau_{1},\, \tau_{2}) = \frac{{\mathcal T}^{(1)}(\vec{x}_{1}, \, \vec{x}_{2}; \tau_{1},\, \tau_{2})}{\sqrt{{\mathcal T}^{(1)}( \tau_{1})}\, \, \sqrt{{\mathcal T}^{(1)}(\tau_{2})}} = \overline{g}^{(1)}(r; \tau_{1},\, \tau_{2})
\nonumber\\
&& = \frac{\int k \, d k j_{0}(k r)\, v_{k}^{*}(\tau_{1}, \tau_{i}) v_{k}(\tau_{2}, \tau_{i}) [\overline{n}_{k}(\tau_{i}) +1 ]  }{\sqrt{\int k \, d k 
|v_{k}(\tau_{1}, \tau_{i})|^2 [\overline{n}_{k}(\tau_{i}) +1 ]} \sqrt{\int k \, d k  |v_{k}(\tau_{2}, \tau_{i})|^2 [\overline{n}_{k}(\tau_{i}) +1 ]}}.
\label{degA16}
\end{eqnarray}
 The degree of first-order coherence computed in the single-polarization approximation, i.e. $\overline{g}^{(1)}(r; \tau_{1},\, \tau_{2})$ coincides with $g^{(1)}(r; \tau_{1},\, \tau_{2})$ because of Eq. (\ref{degA15}). When $\tau_{1} \to \tau_{2}$ and $r\to 0$ we also have that 
\begin{equation}
\lim_{\tau_{1}\to \tau_{2}} g^{(1)}(\vec{x}_{1}, \, \vec{x}_{1}; \tau_{1},\, \tau_{2}) = \lim_{\tau_{1}\to \tau_{2}}\overline{g}^{(1)}(\vec{x}_{2}, \, \vec{x}_{2}; \tau_{1},\, \tau_{2}) = 1.
\label{degA17}
\end{equation}
Thus, in the zero-delay limit and for spatially coincident points the relic gravitons are always first-order coherent. The result implied by Eq. (\ref{degA17}) is actually more general and it holds in all relevant physical regimes, as we shall 
discuss in the following three subsections.

\subsection{Degree of first-order coherence beyond the effective horizon}
For the sake of concreteness we shall first consider the situation where 
the scale factor evolves as a power of the conformal time [i.e. $a(\tau) = (- \tau/\tau_{1})^{-\beta}$ for $\tau < - \tau_{1}$]. The explicit form of $u_{k}(\tau)$ and $v_{k}(\tau)$ follows then from the 
solution of Eqs. (\ref{AA12a}) and (\ref{AA12b}) with the correct boundary conditions 
for $\tau\to - \infty$; the result is given by\footnote{
Equations (\ref{degA18}) and (\ref{degA19}) follow from the solutions of Eqs. (\ref{AA12a}) and (\ref{AA12b}).
The linear combinations $f_{k} = (u_{k} - v_{k}^{*})/\sqrt{2k}$ and $g_{k}= - i (u_{k} + v_{k}^{*})\sqrt{k/2}$
satisfy two decoupled equations that are solved in terms of  Hankel functions of first and second kind \cite{abr1,abr2}.}:
\begin{eqnarray}
u_{k}(\tau) &=& \frac{i}{2} {\mathcal N} \sqrt{- k\tau} \biggl[ H_{\nu+1}^{(1)}(- k\tau) + \biggl( \frac{2 \nu}{ k \tau} - i \biggr)H_{\nu}^{(1)}(- k\tau) \biggr],
\label{degA18}\\
v_{k}(\tau) &=& - \frac{i}{2} {\mathcal N}^{*} \sqrt{- k\tau} \biggl[ H_{\nu+1}^{(2)}(- k\tau) + \biggl( \frac{2 \nu}{ k \tau} - i \biggr)H_{\nu}^{(2)}(- k\tau) \biggr],
\label{degA19}
\end{eqnarray}
where ${\mathcal N}= \sqrt{\pi/2} e^{i \pi(\nu + 1/2)/2}$ and $\nu = (\beta + 1/2)$. 
In the exact de Sitter case case $\beta \to 1$ and 
$\nu \to 3/2$ and the results of Eqs. (\ref{degA18}) and (\ref{degA19}) then imply 
\begin{equation}
u_{k}(\tau) = \biggl(1 - \frac{i}{2 k \tau} \biggr) e^{- i k\tau}, \qquad v_{k}(\tau) = - \frac{i}{ 2 k \tau} e^{i k \tau}.
\label{degA20}
\end{equation}
When the given wavelength is larger than the effective horizon (i.e. 
$k \tau \ll 1$) Eq. (\ref{degA20}) implies that $u_{k}(\tau) \simeq v_{k}(\tau) 
\simeq  i \,(-2 \,k \,\tau)^{-1}$.  The squeezing parameters and the squeezing phases can be determined 
directly from Eqs. (\ref{BB8a}) and (\ref{BB8b}) but the same 
result also follows from the explicit expressions of $u_{k}(\tau)$ and $v_{k}(\tau)$ with the help of Eq. (\ref{degA9}). For instance in the case of an exact de Sitter background of Eq. (\ref{degA20}) the squeezing parameter and the two corresponding phases are: 
\begin{eqnarray}
r_{k} &=& \mathrm{arcsinh}{y} \to - \ln{( - k \tau)} + k^2 \tau^2 + {\mathcal O}(k^4\tau^4),
\label{degA21a}\\
 \delta_{k} &=& k \tau - \arctan{y} \to - k\tau - \frac{\pi}{2} + \frac{8}{3}k^3 \tau^3 + 
 {\mathcal O}(k^5\tau^5),
\label{degA21b}\\
\theta_{k} &=&\frac{\pi}{2} + \arctan{y}  \to 2 k \tau + \pi - \frac{8}{3}k^3 \tau^3 + 
 {\mathcal O}(k^5\tau^5),
\label{degA21c}
\end{eqnarray}
where $y = {\mathcal H}/(2 k) = 1/(- 2 k \tau)$. 
The conformal time coordinate $\tau$ is negative during an exact 
de Sitter phase so that $y\to - \infty$ for typical wavelengths larger than the effective horizon.
In the same limit it also follows from  Eqs. (\ref{degA21b}) and (\ref{degA21c}) that the combination  $(\delta_{k} + \theta_{k}/2)$ is practically vanishing to an excellent approximation:
\begin{equation}
\lim_{|k\tau|\ll1} \biggl( \delta_{k} + \frac{\theta_{k}}{2}\biggr) \to  \frac{4}{3} k^3 \tau^3 +
 {\mathcal O}(k^5 \tau^5) \ll 1.
\label{degA220}
\end{equation}
While the result of Eq. (\ref{degA220}) holds in the exact de Sitter case, in  the quasi-de Sitter case we have instead\footnote{Note that  $\epsilon = [{\mathcal H}^2 - {\mathcal H}^{\prime}]/{\mathcal H}^2$ is the slow-roll parameter written in terms of the conformal time parametrization consistently employed in the present paper. }
 $\nu = (3 -\epsilon)/[2 ( 1- \epsilon)]$. 
In the quasi-de Sitter case the results of Eq. (\ref{degA19}) can then be expanded by recalling that for the range of parameters characteristic of the slow-roll dynamics (i.e. $\epsilon<1$)  we have 
\begin{equation}
H_{\nu}^{(2,\,1)}(z) = \pm \frac{i}{\pi} \biggl(\frac{z}{2}\biggr)^{- \nu} \biggl[ \Gamma(\nu) + \biggl(\frac{z}{2} \biggr)^2 \Gamma(\nu-1) + {\mathcal O}(z^4)\biggr],
\label{degA22}
\end{equation}
where the plus and the minus apply to the Hankel functions of second and first kind 
respectively \cite{abr1,abr2}. Since to leading order in $|k \tau|$ there is is a cancellation  in $v_{k}(\tau)$ and $u_{k}(\tau)$ the correct  asymptotic result follows from Eq. (\ref{degA22})
by keeping the  next-to-leading correction in $|k\tau|$:
\begin{eqnarray}
u_{k}(\tau) &=& \frac{e^{i \pi (\nu - 3/2)/2}}{2\sqrt{2\pi}} \Gamma(\nu)   ( - k \tau)^{- \nu + 1/2}\biggl( i + \frac{x}{2}\biggr), 
\nonumber\\ 
v_{k}(\tau) &=&  \frac{e^{-i \pi (\nu - 3/2)/2}}{2\sqrt{2\pi}} \Gamma(\nu)   ( - k \tau)^{- \nu + 1/2}\biggl( i + \frac{x}{2}\biggr),
\label{degA23}
\end{eqnarray}
which coincides with Eq. (\ref{degA20}) for $\nu= 3/2$ and in the limit $|k\tau|\ll 1$. According to Eq. (\ref{degA23}) the asymptotic values of the phases appearing in Eqs. (\ref{degA21b}) and (\ref{degA21c}) are given by $\delta_{k} = \pi (1/2 - \nu)/2 + {\mathcal O}(k\tau)$ and 
by $(\theta_{k}+ \delta_{k})= \pi (5/2 -\nu)/2 + {\mathcal O}(k\tau)$. Conversely in the quasi-de Sitter case the combination $ (\delta_{k} + \theta_{k}/2)$ of Eq. (\ref{degA220}) is much larger than
${\mathcal O}(k^3 \tau^3)$, i.e. 
\begin{equation}
\lim_{|k\tau|\ll 1} \biggl(\delta_{k} + \frac{\theta_{k}}{2}\biggr) = \frac{\pi (3 - 2\nu)}{4} + {\mathcal O}(k\tau) \simeq - \pi \epsilon/2 + {\mathcal O}(\epsilon^2) + {\mathcal O}(k\tau).
\label{degA24a}
\end{equation}
From the results obtained so far it then follows that the degrees of first-order coherence
of Eq. (\ref{corrT10a}) can be explicitly computed from Eq. (\ref{degA23}) when the relevant wavelengths are larger than the effective horizon:
\begin{eqnarray}
g^{(1)}(r, \tau_{1}, \tau_{2}) &=&   \frac{\int k \,d k \,j_{0}(k r)\, | k^2 \tau_{1} \tau_{2}|^{1/2- \nu} \,[ \overline{n}_{k}(\tau_{i})+1]}{\int k d k | k^2 \tau_{1} \tau_{2}|^{1/2- \nu} [ \overline{n}_{k}(\tau_{i})+1]}
\nonumber\\
&=& \frac{\int k^{2 (1-\nu)} d k j_{0}(k r) [ \overline{n}_{k}(\tau_{i})+1]}{\int k^{2 (1-\nu)} \,d k \, [ \overline{n}_{k}(\tau_{i})+1]} \equiv \overline{g}^{(1)}(r, \tau_{1}, \tau_{2}).
\label{degA24}
\end{eqnarray}
Equation (\ref{degA24}) shows that the degrees of first-order coherence defined in Eq. (\ref{corrT10a}) coincide and the single-polarization approximation gives the same result of the exact Glauber correlator. Since the dependence on $\tau_{1}$ and $\tau_{2}$ disappears from Eq. (\ref{degA24}) the degree of first-order coherence goes always to $1$  beyond the effective horizon. Equation (\ref{degA24}) explains 
and justifies the analog result already mentioned in the zero-delay limit (see Eq. (\ref{degA17})).

\subsection{Degree of first-order coherence inside the effective horizon}
When the expansion rate exceeds the wavenumber a mode is said to be beyond the effective horizon: this does not necessarily have anything to 
do with causality \cite{SW1}.  The qualitative description of the evolution of the tensor modes of the geometry 
stipulates that a given wavelength exits the effective horizon 
(also dubbed sometimes Hubble radius) at a typical conformal time $\tau_{ex}$ (for instance during 
an inflationary stage of expansion) and approximately reenters at $\tau_{re}$, when the 
Universe still expands but at a decelerated rate. Inside the effective horizon, i.e. in the limit $|k \tau \gg 1$, Eqs. (\ref{degA21a})--(\ref{degA21c}) the squeezing parameters become 
\begin{equation}
r_{k} \ll 1, \qquad \delta_{k} \simeq k \tau, \qquad \theta_{k} \simeq \pi/2, \qquad \delta_{k} + \frac{\theta_{k}}{2} \simeq k\tau + \frac{\pi}{4} \gg 1.
\label{lim}
\end{equation}
By taking the large argument limits of the corresponding Hankel functions in Eqs. (\ref{degA18}) and (\ref{degA19}), $u_{k}(\tau)$ and $v_{k}(\tau)$ can be 
determined when $|k\tau| \gg 1$:
\begin{equation}
u_{k}(\tau) = e^{- i k\tau}\biggl[ 1  + {\mathcal O}\biggl(\frac{1}{k \tau}\biggr)\biggr], \qquad  
v_{k}(\tau) = - e^{i k\tau}\biggl[ 1  + {\mathcal O}\biggl(\frac{1}{k \tau}\biggr)\biggr],
\label{degA250}
\end{equation}
implying $\delta_{k} \simeq k\tau $ and $(\theta_{k} + \delta_{k}) \simeq k\tau$.
These results also imply that $\delta_{k} + \theta_{k}/2 \simeq k\tau \gg 1$ and apply 
when the modes of the fields are inside the effective inflationary horizon. 
Equation (\ref{degA250}) is however not applicable during the radiation of matter phases 
but only describe the modes inside the effective horizon during the inflationary phase. 

To compute the degrees of coherence inside the effective horizon after the end of inflation 
it is simpler to avoid specific exact solution such as the ones 
discussed in Eqs. (\ref{degA18})--(\ref{degA19}) and directly work within an
 appropriate WKB approximation where the evolution of $u_{k}$ and $v_{k}$ will be approximate 
but more generally applicable. The strategy will be to ensure the correct 
normalization of $u_{k}(\tau)$ and $v_{k}(\tau)$ in the limit $k\tau \gg 1$
and then compute their form when the modes reenter either during the 
radiation-dominated phase or during the matter epoch. This analysis has been 
relegated to appendix \ref{APPB} and it can be found in the current literature. 
As a consequence the functions $u_{k}(\tau)$ and $v_{k}(\tau)$ can be expressed as:
\begin{eqnarray}
u_{k}(\tau) &=&  \biggl[\biggl(1 - \frac{i {\mathcal H}}{2 k} \biggr) c_{+}(k) e^{- i k \tau} - \frac{i {\mathcal H}}{2 k} c_{-}(k) e^{i k\tau}\biggr] \to c_{+}(k) e^{- i k\tau} , 
\label{degA25}\\
v_{k}(\tau) &=& \biggl[\frac{i {\mathcal H}}{2 k } c_{+}^{*}(k) e^{i k\tau} - \biggl( 1 - \frac{i {\mathcal H}}{2 k} \biggr) c_{-}^{*}(k) e^{- i k \tau} \biggr] \to - c_{-}^{*}(k) e^{-i k \tau},
\label{degA26}
\end{eqnarray}
where $c_{\pm}(k)$ have been determined in Eq. (\ref{CC4}) from the exact matching across 
the turning points $\tau_{ex}$ and $\tau_{re}$ (see also \cite{GR3,blue1}). As a function of $k \tau$ the explicit expressions of Eqs. (\ref{degA25})--(\ref{degA26}) hold in the limit $k \tau \gg 1$. The values of $c_{\pm}(k)$ depend on the reentry and on the exit of the given mode and 
they can be usefully approximated from Eq. (\ref{CC4}) as\footnote{We correct here a minor typo in the second paper of Ref. \cite{blue1} where $\tau_{ex}$ and $\tau_{re}$ have been interchanged in the phase $\nu_{\pm}$.}
\begin{equation}
c_{\pm}(k) = \frac{ e^{- i \nu_{\pm}(k) }}{2 i}\biggl[ \biggl(\frac{a_{re}}{a_{ex}}\biggr) \biggl( i \mp \frac{{\mathcal H}_{re}}{k} \biggr) \pm (1 + i) \biggl(\frac{a_{ex}}{a_{re}}\biggr) \biggr],
\label{degA27}
\end{equation}
where $\nu_{\pm}(k) = k (\tau_{ex} \mp \tau_{re})$ and 
the relation $|c_{+}(k)|^2 - |c_{-}(k)|^2 =1$ holds explicitly. Equation (\ref{degA27}) 
follows from the results of appendix \ref{APPB} and it implies that, for wavelengths shorter than the effective horizon, the degree of first-order coherence becomes 
\begin{eqnarray}
g^{(1)}(r, \tau_{1}, \tau_{2}) = \overline{g}^{(1)}(r, \tau_{1}, \tau_{2}) = \frac{\int k \, d k\,  j_{0}(k r)\, |c_{-}(k)|^2 \, [\overline{n}_{k}(\tau_{i}) +1] e^{- i k (\tau_{1} - \tau_{2})}}{\int k \, d k\, \, |c_{-}(k)|^2 \, [\overline{n}_{k}(\tau_{i}) +1] }.
\label{degA29}
\end{eqnarray}
The result of Eq. (\ref{degA29}) goes to $1$ in the zero-delay limit; conversely if $\tau_{1}\neq \tau_{2}$ the degree of first-order coherence 
computed from Eq. (\ref{degA29}) is always smaller than $1$, i.e. $|g^{(1)}(0,\tau_{1},\tau_{2})| \leq 1$ where 
the sign of equality holds for $\tau_{1} = \tau_{2}$. 

\subsection{Single-mode approximation}
If the integrals over the modes and the sum over the polarizations are replaced by a single quantum oscillator
the two terms appearing in the exponent of the operator ${\mathcal S}(z)$ become:
\begin{equation}
\sum_{\beta} \int d^{3} k \,\, z_{k\,\,\beta}^{*} \, \hat{a}_{\vec{k}\,\,\beta} 
\hat{a}_{-\vec{k}\,\,\beta} \to \frac{z^{*}}{2} \,\, \hat{a}^{ 2},\qquad  \sum_{\beta} \int d^{3} k \,\, z_{k\,\,\beta} \, \hat{a}^{\dagger}_{-\vec{k}\,\,\beta} 
\hat{a}^{\dagger}_{\vec{k}\,\,\beta} \to \frac{z}{2} \,\, \hat{a}^{\dagger\,\,2}.
\label{degA30}
\end{equation}
This is, in a nutshell, the idea of the single-mode approximation which is not fully realistic 
since, in Eq. (\ref{degA30}) the squeezing parameter and the 
phases must anyway follow the dynamics coming from the full Hamiltonian. With these caveats 
the single-mode approximation is  crude but  interesting, at least for comparison. 
 We shall denote with the calligraphic style the continuous-mode operators while the single-mode operators will be denoted by the standard capital letter in roman style; so for instance 
\begin{equation}
{\mathcal S}(z) \to S(z) = e^{ \frac{z^{*}}{2}\,\hat{a}^2 - \frac{z}{2}\,\hat{a}^{\dagger \,2}},\qquad 
{\mathcal R}(\delta) \to R(\delta) = e^{- \frac{i}{2}\, \delta \,\hat{a}^{\dagger} \hat{a}},
\label{degA31a}
\end{equation}
where $z = r \, e^{i \theta}$, $\alpha= |\alpha| e^{i \varphi}$ and so on and so forth.
Using the properties of Eq. (\ref{degA31a}) we will have, for instance, 
\begin{equation}
R^{\dagger}(\delta) S^{\dagger}(z) \hat{a} S(z) R(\delta) = e^{- i \delta} \cosh{r} \,\,\hat{a} - e^{i (\theta+ \delta)} \sinh{r} \,\,\hat{a}^{\dagger}. 
\label{degA31b}
\end{equation}
The degree of first-order coherence in the single-mode case will then be given by 
\begin{equation}
g^{(1)}_{s}(\tau_{1}, \tau_{2}) = \frac{\langle\hat{a}^{\dagger}(\tau_{1}) \hat{a}(\tau_{2}) \rangle}{\sqrt{\langle\hat{a}^{\dagger}(\tau_{1}) \hat{a}(\tau_{1}) \rangle} \, \sqrt{\langle\hat{a}^{\dagger}(\tau_{2}) \hat{a}(\tau_{2}) \rangle}},
\label{degA32}
\end{equation}
where the subscript $s$ reminds that we are here discussing the single-mode approximation. 
In spite of the statistical properties of the state $g_{s}^{(1)}\to 1$ in the zero-delay limit\footnote{The second-order correlations in the single-mode approximation will not be that trivial but they will instead depend on the 
specific properties of the quantum state. } (i.e. for  $\tau_{1} \to \tau_{2}$).  The single-mode limit can also be implemented in a slightly different way by introducing two different oscillators
$[\hat{c}, \hat{c}^{\dagger}] =1$ and $[\hat{d}, \hat{d}^{\dagger}] = 1$ (with $[\hat{c}, \hat{d}]=0$).
In this case the two-mode squeezing and rotation operators \cite{CV1,CV2}  (analogs of $R(\delta)$ and $S(z)$ but valid in the two-mode case) imply 
\begin{eqnarray}
\overline{R}^{\dagger}(\delta) \,\overline{S}^{\dagger}(z) \,\hat{c} \, \overline{S}(z) \,\overline{R}(\delta) &=& e^{- i \delta} \cosh{r} \hat{c} - e^{i (\theta+ \delta)} \sinh{r} \hat{d}^{\dagger},
\label{degA33a}\\
\overline{R}^{\dagger}(\delta) \,  \overline{S}^{\dagger}(z) \,\hat{d} \,\overline{S}(z) \,\overline{R}(\delta) &=& e^{- i \delta} \cosh{r} \hat{d} - e^{i (\theta+ \delta)} \sinh{r} \hat{c}^{\dagger}.
\label{degA33b}
\end{eqnarray}
In this case the sum $\hat{a} = (\hat{c} + \hat{d})/\sqrt{2}$ obeys $[\hat{a}, \hat{a}^{\dagger}] =1$.
By summing Eqs. (\ref{degA33a}) and (\ref{degA33b}), Eq. (\ref{degA31b}) 
is recovered. The degree of first-order coherence will then follow, with the same properties, from Eq. (\ref{degA32}).
In the approach of Eqs. (\ref{degA33a}) and (\ref{degA33b}) the multiplicity of a given state 
is the sum of the individual multiplicities, i.e. $\langle z| \hat{a}^{\dagger} \hat{a} | z\rangle= 
(\langle z| \hat{c}^{\dagger} \hat{c} | z\rangle + \langle z| \hat{d}^{\dagger} \hat{d} | z\rangle)/2 = 
\sinh^2{r}$. 

\renewcommand{\theequation}{4.\arabic{equation}}
\setcounter{equation}{0}
\section{Second-order coherence of relic gravitons}
\label{sec4}
\subsection{General expressions for the degree of second-order coherence}
The  Hanbury Brown-Twiss correlations preliminary presented in 
 Eq. (\ref{corrT9a}) will now be estimated. For the sake of convenience the 
 discussion mirrors exactly the same steps of the previous section. After 
 determining the correlation functions, the degrees of second-order coherence will be explicitly 
 discussed in various limits. From the expressions of $\hat{\mu}^{(-)}(x)$ and $\hat{\mu}^{(+)}(x)$ 
Eq. (\ref{corrT9a}) becomes:
\begin{eqnarray}
 {\mathcal T}^{(2)}(x_{1}, x_{2}) &=& \frac{1}{(2\pi)^6} \int \frac{d^{3} k_{1}}{\sqrt{2 k_{1}}} 
  \int \frac{d^{3} k_{2}}{\sqrt{2 k_{2}}}  \int \frac{d^{3} k_{3}}{\sqrt{2 k_{3}}}  \int \frac{d^{3} k_{4}}{\sqrt{2 k_{4}}} 
 e^{- i (\vec{k}_{1} + \vec{k}_{4})\cdot\vec{x}_{1}}  e^{- i (\vec{k}_{2} + \vec{k}_{3})\cdot\vec{x}_{2}} 
\nonumber\\
&\times& \sum_{\alpha_{1}} \, \sum_{\alpha_{2}}  \, \sum_{\alpha_{3}}  \,\sum_{\alpha_{4}}  \,\,e_{ij}^{(\alpha_{1})}(\hat{k}_{1}) \,\, e_{kl}^{(\alpha_{2})}(\hat{k}_{2}) \,\, e_{kl}^{(\alpha_{3})}(\hat{k}_{3}) \,\,
e_{ij}^{(\alpha_{4})}(\hat{k}_{4}) 
\nonumber\\
&\times& \langle \hat{a}^{\dagger}_{- \vec{k}_1, \, \alpha_{1}}(\tau_{1}) \,\, \hat{a}^{\dagger}_{- \vec{k}_2, \, \alpha_{2}}(\tau_{2}) \,\,\hat{a}_{ \vec{k}_3, \, \alpha_{3}}(\tau_{2})  \,\,\hat{a}_{ \vec{k}_4, \, \alpha_{4}}(\tau_{1})  \rangle.
\label{degB1}
\end{eqnarray}
The expectation value appearing in the last line of 
Eq. (\ref{degB1}) must be first referred to the initial time $\tau_{i}$ when, by definition, all 
the relevant modes are inside the effective horizon (i.e. $k \tau_{i} \gg 1$).  For this purpose, using Eqs. (\ref{degA4}) and (\ref{degA5}) we can write
\begin{eqnarray}
&& \langle \hat{a}^{\dagger}_{- \vec{k}_1, \, \alpha_{1}}(\tau_{1}) \,\, \hat{a}^{\dagger}_{- \vec{k}_2, \, \alpha_{2}}(\tau_{2}) \,\,\hat{a}_{ \vec{k}_3, \, \alpha_{3}}(\tau_{2})  \,\,\hat{a}_{ \vec{k}_4, \, \alpha_{4}}(\tau_{1})  \rangle
\nonumber\\
&&= v_{k_{1},\,\alpha_{1}}^{*}(\tau_{1},\tau_{i}) v_{k_{2},\,\alpha_{2}}^{*}(\tau_{2}, \tau_{i}) v_{k_{3},\,\alpha_{3}}(\tau_{2},\tau_{i}) v_{k_{4},\,\alpha_{4}}(\tau_{1},\tau_{i})
\langle \hat{b}_{\vec{k}_{1},\,\alpha_{1}} \hat{b}_{\vec{k}_{2}, \alpha_{2}}
 \hat{b}_{-\vec{k}_{3},\,\alpha_{3}}^{\dagger} \hat{b}_{-\vec{k}_{4},\,\alpha_{4}}^{\dagger}\rangle
 \nonumber\\
 &&+  v_{k_{1},\,\alpha_{1}}^{*}(\tau_{1}, \tau_{i}) u_{k_{2},\,\alpha_{2}}^{*}(\tau_{2},\tau_{i}) u_{k_{3},\,\alpha_{3}}(\tau_{2},\tau_{i}) v_{k_{4},\,\alpha_{4}}(\tau_{1},\tau_{i})
\langle \hat{b}_{\vec{k}_{1},\,\alpha_{1}} \hat{b}_{\vec{k}_{2}, \alpha_{2}}^{\dagger}
 \hat{b}_{-\vec{k}_{3}, \alpha_{3}} \hat{b}_{-\vec{k}_{4}, \alpha_{4}}^{\dagger}\rangle.
 \label{degB2}
 \end{eqnarray}
Since the relic graviton background  is not polarized, as in Eq. (\ref{degA11}) the
standard quantum-mechanical initial conditions imply $u_{k,\oplus}(\tau,\tau_{i}) = u_{k\,\otimes}(\tau,\tau_{i}) = u_{k}(\tau,\tau_{i})$ 
 and $v_{k,\oplus}(\tau,\tau_{i}) = v_{k\,\otimes}(\tau,\tau_{i}) = v_{k}(\tau,\tau_{i})$.
 If the initial state is not the vacuum\footnote{ As already mentioned in section \ref{sec3} 
this parametrization of the initial state will be complemented by the considerations 
of  section \ref{sec5} where a more specific discussion will be outlined.}
 at $\tau_{i}$ the expectation value can be expressed as $\langle  \hat{b}^{\dagger}_{\vec{k}, \, \alpha}(\tau_{i}) \hat{b}_{\vec{p}, \, \beta}(\tau_{i}) \rangle = \overline{n}_{k}(\tau_{i}) \,\delta_{\alpha\beta} \, \delta^{(3)}(\vec{k}- \vec{p})$ where $\overline{n}_{k}(\tau_{i})$ is the average multiplicity of the initial state and the standard case vacuum result will be recovered from the final HBT correlations in the limit $\overline{n}_{k}(\tau_{i}) \to 0$.
After inserting into Eq. (\ref{degB1})  the results  of Eq. (\ref{degB2}),  the explicit form of the HBT 
correlations is: 
 \begin{eqnarray}
 {\mathcal T}^{(2)}(x_{1}, x_{2}) &=& \frac{1}{(2\pi)^6} \int \frac{d^{3} k}{k} \, \int \frac{d^{3} p}{p}\,[\overline{n}_{k}(\tau_{i}) + 1]
 \,[\overline{n}_{p}(\tau_{i}) + 1] \biggl\{
 4 |v_{k}(\tau_{1},\tau_{i})|^2 \,\, |v_{p}(\tau_{2}, \tau_{i})|^2 
\nonumber\\
&+& \frac{1}{4} [ 1 + (\hat{k}\cdot\hat{p})^2] [ 1 + 3(\hat{k}\cdot\hat{p})^2] \biggl[ v_{k}^{*}(\tau_{1},\tau_{i}) 
 v_{p}^{*}(\tau_{2},\tau_{i}) v_{k}(\tau_{2},\tau_{i})v_{p}(\tau_{1},\tau_{i}) 
 \nonumber\\
 &+& v_{k}^{*}(\tau_{1},\tau_{i}) u_{k}^{*}(\tau_{2},\tau_{i}) u_{p}(\tau_{2},\tau_{i})v_{p}(\tau_{1},\tau_{i})\biggr] 
 e^{- i (\vec{k} - \vec{p})\cdot\vec{r}}
 \biggr\}.
 \label{degB3}
 \end{eqnarray}
Equation (\ref{degB3}) differs from the single-polarization approximation
described by Eqs. (\ref{corrT8b}) and (\ref{corrT9b}). In the single-polarization 
approximation Eq. (\ref{degB3}) must then be replaced by the result 
that follows, in the same physical situation, from Eqs. (\ref{corrT8b}) and (\ref{corrT9b}):
 \begin{eqnarray}
 {\mathcal S}^{(2)}(x_{1}, x_{2}) &=& \frac{1}{4(2\pi)^6} \int \frac{d^{3} k}{k} \, \int \frac{d^{3} p}{p} 
 [\overline{n}_{k}(\tau_{i}) + 1]
 \,[\overline{n}_{p}(\tau_{i}) + 1]
\nonumber\\
&\times& \biggl\{ |v_{k}(\tau_{1},\tau_{i})|^2 \,\, |v_{p}(\tau_{2}, \tau_{i})|^2 
 +\biggl[ v_{k}^{*}(\tau_{1},\tau_{i}) 
 v_{p}^{*}(\tau_{2},\tau_{i}) v_{k}(\tau_{2},\tau_{i})v_{p}(\tau_{1},\tau_{i}) 
 \nonumber\\
 &+& v_{k}^{*}(\tau_{1},\tau_{i}) 
 u_{k}^{*}(\tau_{2},\tau_{i}) u_{p}(\tau_{2},\tau_{i})v_{p}(\tau_{1},\tau_{i})\biggr] 
 e^{- i (\vec{k} - \vec{p})\cdot\vec{r}}\biggr\}.
 \label{degB4}
\end{eqnarray}
Following the same logic, the explicit expression given in Eq. (\ref{degB4}) together 
with Eq. (\ref{degA14}) shall be inserted into the second expression of Eq. (\ref{corrT11a}) and this will
lead to the explicit expression of the degree of second-order coherence 
in the single-polarization approximation.  The results of Eqs. 
(\ref{degB3}) and (\ref{degB4})  imply that the degrees of second-order coherence of Eq. (\ref{corrT11a}) 
 only depend on $r= |\vec{x}_{1} - \vec{x}_{2}|$:
\begin{eqnarray} 
g^{(2)}(r, \tau_{1}, \tau_{2}) &=& \frac{{\mathcal T}^{(2)}(r, \tau_{1}, \tau_{2})}{{\mathcal T}^{(1)}(\tau_{1})\, {\mathcal T}^{(1)}(\tau_{2})}, \qquad {\mathcal T}^{(1)}(\tau) = \frac{1}{\pi^2} \int k \,dk |v_{k}(\tau)|^2 [\overline{n}_{k}(\tau_{i}) +1],
\label{degB5}\\
\overline{g}^{(2)}(r, \tau_{1}, \tau_{2}) &=& \frac{{\mathcal S}^{(2)}(r, \tau_{1}, \tau_{2})}{{\mathcal S}^{(1)}(\tau_{1})\, {\mathcal S}^{(1)}(\tau_{2})}, \qquad {\mathcal S}^{(1)}(\tau) = \frac{1}{4\pi^2} \int k \,dk |v_{k}(\tau)|^2 [\overline{n}_{k}(\tau_{i}) +1].
\label{degB6}
\end{eqnarray}
\subsection{Degree of second-order coherence beyond the effective horizon}
When the wavelengths of the gravitons exceed the effective horizon, 
the functions $u_{k}(\tau, \tau_{i})$ and $v_{k}(\tau,\tau_{i})$ are given by  
the result of Eqs. (\ref{degA23}). Using Eq. (\ref{degB3}),
 to leading order in $k \tau_{1} \ll 1$ and $p \tau_{2} \ll 1$ 
the HBT correlations become:
\begin{eqnarray}
{\mathcal T}^{(2)}(r,\tau_{1}, \tau_{2}) &=& \frac{2^{4 \nu} \, \Gamma^4(\nu)}{4096 \, \pi^{8}} \, \int k \,d k \,[\overline{n}_{k}(\tau_{i}) +1]
 |k \tau_{1}|^{ 1 - 2 \nu} \, \int p \,d p\, [\overline{n}_{p}(\tau_{i}) +1] |p \tau_{2}|^{1 - 2 \nu}
 \nonumber\\
 &\times& {\mathcal E}(k, p, r) \biggl[ 1 + {\mathcal O}(k^2 \tau_{1}^2)\biggr] \biggl[ 1 + {\mathcal O}(p^2 \tau_{2}^2)\biggr],
\label{degB7}
\end{eqnarray} 
where the function ${\mathcal E}(k, p, r)$ appearing in Eq. (\ref{degB7}) can be expressed as:
\begin{eqnarray}
{\mathcal E}(k, p, r) &=& \int d \hat{k} \, \int d\hat{p} \biggl\{ 1 + \frac{1}{8} [ 1 + (\hat{k}\cdot\hat{p})^2 ] [ 1 + 3(\hat{k}\cdot\hat{p})^2 ] e^{- i (\vec{k} -\vec{p})\cdot\vec{r}}\biggr\}
\nonumber\\
&=&  \frac{704 \pi^2}{15} - \frac{352 \pi^2}{45}( k^2 r^2 + p^2 r^2)  + {\mathcal O}( k^4 r^4) + {\mathcal O}(p^4 r^4) +
{\mathcal O}(k^2 p^2 r^4).
\label{degB8}
\end{eqnarray}
In Eq. (\ref{degB8}) $d \hat{k}$ and  $ d \hat{p}$ denote the angular integrations 
over the directions of the comoving three-momenta. The result of Eqs. (\ref{degB7})--(\ref{degB8})
can be compared with the intensity correlations computed in the single-polarization approximation. For this purpose  Eq. (\ref{degA23}) must be inserted into Eq. (\ref{degB4}) so that, to leading order in $k \tau_{1} \ll 1$ and $p\tau_{2} \ll 1$ the intensity correlations in the single-polarization approximation will be
 \begin{eqnarray}
{\mathcal S}^{(2)}(r,\tau_{1}, \tau_{2}) &=& \frac{2^{4 \nu} \, \Gamma^4(\nu)}{256  (2 \pi)^{8}} \, \int k \,d k\, [\overline{n}_{k}(\tau_{i}) +1]\,|k \,\tau_{1}|^{ 1 - 2 \nu} \, \int p \,d p\, [\overline{n}_{p}(\tau_{i}) +1] \, |p \,\tau_{2}|^{1 - 2 \nu}
 \nonumber\\
 &\times& \overline{{\mathcal E}}(k, p, r) \biggl[ 1 + {\mathcal O}(k^2 \tau_{1}^2)\biggr] \biggl[ 1 + {\mathcal O}(p^2 \tau_{2}^2)\biggr],
\label{degB9}
\end{eqnarray} 
where this time the analog of Eq. (\ref{degB8}) is given, with the same notations, by:
\begin{eqnarray}
\overline{{\mathcal E}}(k, p, r) &=& \int d \hat{k} \, \int d\hat{p} \biggl[ 1 + 2 e^{- i (\vec{k} - \vec{p})\cdot\vec{r}} \biggr] 
\nonumber\\
&=& 48 \pi^2 - \frac{16}{3} \pi^2 ( k^2 r^2 + p^2 r^2) + {\mathcal O}( k^4 r^4) + {\mathcal O}(p^4 r^4) +
{\mathcal O}(k^2 p^2 r^4).
\label{degB10}
\end{eqnarray}
The degrees of second-order 
coherence defined in Eqs. (\ref{degB5}) and (\ref{degB6}) follow then from Eqs. 
(\ref{degB8}) and (\ref{degB9}) 
by recalling the explicit form of ${\mathcal T}^{(1)}(\tau)$ and 
${\mathcal S}^{(1)}(\tau)$ given in Eqs. (\ref{degA14}) and (\ref{degA15}). 
Bearing in mind the results for the first-order correlations, it turns out that 
the intensity correlations are factorized as follows:
\begin{eqnarray} 
{\mathcal T}^{(2)}(r,\tau_{1}, \tau_{2}) &\simeq& \frac{41}{30}\, {\mathcal T}^{(1)}(\tau_{1}) {\mathcal T}^{(1)}(\tau_{2}),
\label{degB11}\\
{\mathcal S}^{(2)}(r,\tau_{1}, \tau_{2}) &\simeq& 3 \,{\mathcal S}^{(1)}(\tau_{1}) {\mathcal S}^{(1)}(\tau_{2}),
\label{degB12}
\end{eqnarray}
when the wavelengths exceed the effective horizon.  The approximate equalities remind that Eqs. (\ref{degB11})--(\ref{degB12}) hold in the limits $| k \tau_{1}| \ll1 $ and $|p \tau_{2}| \ll 1$.
Thanks to Eqs. (\ref{degB5}) and (\ref{degB6}) the results of Eqs. (\ref{degB11}) and (\ref{degB12}) imply that\footnote{Note that a previous analysis of  $g^{(2)}(r, \tau_{1}, \tau_{2})$ in the limits $|k\tau_{1}| \ll 1$ and $|k \tau_{2}| \ll 1$ led to 
$71/60\simeq 1.18$ \cite{GR6} while this more accurate analysis  shows that this result must be corrected as $41/30 \simeq 1.36$.}
\begin{equation}
g^{(2)}(r, \tau_{1}, \tau_{2}) \to \frac{41}{30},\qquad \overline{g}^{(2)}(r, \tau_{1}, \tau_{2}) \to 3.
\label{degB13}
\end{equation}
According to Eq. (\ref{degB13})  the degree of second-order coherence is always super-Poissonian when the relevant wavelengths exceed the effective horizon since both $g^{(2)}$ and $\overline{g}^{(2)}$ are larger than $1$. In the single-polarization approximation  the degree of second-order coherence goes to $3$ while the presence of the polarization reduces the degree of second-order coherence. 

\subsection{Degree of second-order coherence inside the effective horizon}
The initial conditions of the Einstein-Boltzmann hierarchy (both for the scalar and for the 
tensor modes of the geometry) are set before matter-radiation equality when the relevant 
wavelengths are larger than the Hubble radius at the corresponding epoch \cite{HBT3,HBT4}. 
For an experiment probing the degree of second-order coherence in the CMB,
the results of Eq. (\ref{degB13}) are the most relevant ones. If we are however interested 
in gravitons whose frequencies are comparable with the operating window of wide-band interferometers 
(i.e. between few Hz and $10$ kHz),  the relevant expression of the degree of second-order coherence 
follows when the wavelengths of the gravitons are all within the effective horizon. As already established in Eqs. (\ref{degA25})--(\ref{degA26}), inside the effective horizon, 
 $u_{k}(\tau) = c_{+}(k) e^{- i k\tau}$ and $v_{k}(\tau) = - c_{-}(k)^{*} e^{i k \tau}$. In 
 the limit $k \tau_{1} \gg 1$ and $p \tau_{2} \gg 1$, Eq. (\ref{degB3}) reads
\begin{eqnarray}
{\mathcal T}^{(2)}(r,\tau_{1}, \tau_{2}) &=& \frac{1}{(2\pi)^6} \int \frac{d^{3} k}{k} [\overline{n}_{k}(\tau_{i}) +1] \int \frac{d^3 p}{p} [\overline{n}_{p}(\tau_{i}) +1] \,\biggl\{ 4 |c_{-}(k)|^2 \, 
|c_{-}(p)|^2 
\nonumber\\
&+& \frac{1}{4} [ 1 + (\hat{k}\cdot\hat{p})^2] [ 1 + 3 (\hat{k}\cdot\hat{p})^2]  \biggl[ |c_{-}(k)|^2 |c_{-}(p)|^2 
\nonumber\\
&+& c_{-}(k) c_{+}^{*}(k) c_{+}(p) c_{-}^{*}(p) \biggr] e^{- i (k -p)(\tau_{1} - \tau_{2})} e^{- i (\vec{k} - \vec{p})\cdot\vec{r}}\biggr\}.
\label{degB15}
\end{eqnarray}
The last to terms of Eq. (\ref{degB15}) can be rewritten by  factoring  $|c_{-}(k)|^2 \, 
|c_{-}(p)|^2 $ and by using Eqs. (\ref{degA27}) in the obtained expression; the result 
of this step is given by:
\begin{equation}
 |c_{-}(k)|^2 |c_{-}(p)|^2 \biggl[ 1 
+ \frac{c_{+}^{*}(k) c_{+}(p)}{ c^{*}_{-}(k) c_{-}(p)} \biggr] \simeq |c_{-}(k)|^2 |c_{-}(p)|^2 
\biggl\{ 1 + \biggl[\frac{( i - {\mathcal H}_{re}/k)}{( i + {\mathcal H}_{re}/k)} \biggr] 
\biggl[\frac{( i + {\mathcal H}_{re}/p)}{( i - {\mathcal H}_{re}/p)} \biggr]\biggr\}.
\label{degB16}
\end{equation}
Equation (\ref{degB16}) can be explicitly estimated in two complementary 
limits. In the first case $k \tau_{re} = p \tau_{re} \simeq 1$; this choice corresponds 
to the situation where, in the vicinity of the turning points, $|a^{\prime\prime}/a| \neq 0$.
If the modes reenter when the condition $|a^{\prime\prime}/a| \to 0$
then $k\tau_{re} <1$ and $p\tau_{re} < 1 $ (see also appendix \ref{APPB} and discussion therein). 
In both situations the results are similar and Eq. (\ref{degB16}) can be approximated as:
\begin{eqnarray}
{\mathcal T}^{(2)}(r,\tau_{1}, \tau_{2}) &=& \frac{4}{(2\pi)^6} \int \frac{d^{3} k}{k} [\overline{n}_{k}(\tau_{i}) +1] \int \frac{d^3 p}{p} [\overline{n}_{p}(\tau_{i}) +1] \,\biggl\{ 4 |c_{-}(k)|^2 \, 
|c_{-}(p)|^2 
\nonumber\\
&+& \frac{1}{8} [ 1 + (\hat{k}\cdot\hat{p})^2] [ 1 + 3 (\hat{k}\cdot\hat{p})^2]  \biggl[ |c_{-}(k)|^2 |c_{-}(p)|^2 
\nonumber\\
&+& c_{-}(k) c_{+}^{*}(k) c_{+}(p) c_{-}^{*}(p) \biggr] e^{- i (k -p)(\tau_{1} - \tau_{2})} e^{- i (\vec{k} - \vec{p})\cdot\vec{r}}\biggr\}.
\label{degB17}
\end{eqnarray}
The result of the angular integration appearing in Eq. (\ref{degB17}) is a complicated 
function of $k r$ and $p r$ that goes to a constant for $k r < 1$ and $p r < 1$. Therefore the degrees of 
second-order coherence will receive the dominant contribution for $k r \sim p r \sim {\mathcal O}(1)$:
\begin{eqnarray}
&& g^{(2)}( \tau_{1}, \tau_{2}) \simeq \frac{41}{30} \frac{\int k\,d k\,[\overline{n}_{k}(\tau_{i}) +1]\,
|c_{-}(k)|^2 \int p\,d p \,|c_{-}(p)|^2 [\overline{n}_{p}(\tau_{i}) +1] e^{- i (k -p)\Delta\tau}}{\int k \,d k  [\overline{n}_{k}(\tau_{i}) +1] 
|c_{-}(k)|^2 \int p \,d p \,|c_{-}(p)|^2 [\overline{n}_{p}(\tau_{i}) +1] },
\label{degB18}\\
&& \overline{g}^{(2)}(\tau_{1}, \tau_{2}) \simeq 3 \frac{\int k \, d k  [\overline{n}_{k}(\tau_{i}) +1] 
|c_{-}(k)|^2 \int p \,d p \,|c_{-}(p)|^2 [\overline{n}_{p}(\tau_{i}) +1] e^{- i (k -p)\Delta\tau}}{\int k \,d k\,[\overline{n}_{k}(\tau_{i}) +1] 
|c_{-}(k)|^2 \int p \, d p \,|c_{-}(p)|^2 [\overline{n}_{p}(\tau_{i}) +1]},
\label{degB19}
\end{eqnarray}
where  $\Delta\tau = \tau_{1}- \tau_{2}$. 
Equations (\ref{degB18}) and (\ref{degB19}) coincide with Eq. (\ref{degB13}) 
in the zero-delay limit. When $\tau_{1} \neq \tau_{2}$ it can be demonstrated that $|g^{(2)}(\tau) |< g^{(2)}(0)$ and  $|\overline{g}^{(2)}(\tau)| < \overline{g}^{(2)}(0)$ which implies, in a quantum optical language, that the degree of second-order coherence is not only super-Poissonian but also bunched \cite{QO1,loudon}. 

\subsection{Single-mode approximation and its physical interpretation}
The degree of first-order coherence has been analyzed in the single-mode 
approximation at the end of section \ref{sec3} and in full analogy 
with the definition of Eq. (\ref{degA32}) the degree of
 second-order (temporal) coherence will be defined as:
\begin{equation}
g_{s}^{(2)}(\tau_{1}, \tau_{2}) = \frac{\langle \hat{a}^{\dagger}(\tau_{1})  \hat{a}^{\dagger}(\tau_{2}) \, \hat{a}(\tau_{2})\, \hat{a}(\tau_{1})\rangle}{\langle \hat{a}^{\dagger}(\tau_{1}) \, \hat{a}(\tau_{1})\rangle \langle \hat{a}^{\dagger}(\tau_{2}) \, \hat{a}(\tau_{2})\rangle}.
\label{degC1}
\end{equation}
In the zero-delay limit Eq. (\ref{degC1}) becomes: 
\begin{equation}
\lim_{\tau_{1} \to \tau_{2}} g_{s}^{(2)}(\tau_{1},\tau_{2}) = g_{s}^{(2)}= \frac{\langle \hat{a}^{\dagger} \hat{a}^{\dagger}\, \hat{a}\, \hat{a}\rangle}{\langle \hat{a}\, \hat{a}\rangle^2},
\label{degC2}
\end{equation}
where, for simplicity, we will employ the notation $g^{(2)}_{s} = g^{(2)}_{s}(0)$. After using the commutation relations Eq. (\ref{degC2}) can be expressed in terms of $\hat{N} = \hat{a}^{\dagger} \hat{a}$ and of the 
dispersion $\sigma^2$:
\begin{equation}
g_{s}^{(2)} = \frac{\sigma^2 - \langle \hat{N} \rangle + \langle \hat{N} \rangle^2}{\langle \hat{N} \rangle^2}, \qquad \sigma^2 = \langle \hat{N}^2 \rangle - \langle \hat{N} \rangle^2.
\label{degC3}
\end{equation}
 Equation (\ref{degC3}) 
is often presented in terms of the so-called Mandel $Q$ parameter \cite{QO1} defined, within our notations, as :
$Q= \langle \hat{N} \rangle [ g_{s}^{(2)} -1]$ implying  $Q = \sigma^2/\langle \hat{N} \rangle - 1$.
In the case of a single-mode squeezed state we have that the previous quantities can be all expressed 
in terms of a single parameter which is the average multiplicity of the state denoted hereunder by $\langle \hat{N} \rangle = \overline{n}_{sq} = \sinh^2 r$:
\begin{equation}
g^{(2)}_{s} = 3 + \frac{1}{\overline{n}}, \qquad Q = ( 2 \overline{n} +1). 
\label{degC5}
\end{equation}
In the case of a coherent state the Mandel parameter vanishes 
so that the result of Eq. (\ref{degC5}) can be dubbed by saying that 
the degree of second-order coherence is super-Poissonian. 
The results obtained in the present section suggest the following hierarchy:
\begin{equation}
g_{s}^{(2)} = \overline{g}^{(2)} > g^{(2)} > 1.
\label{degC5a}
\end{equation}
The first equality follows from the comparison between the single-mode approximation 
of Eqs. (\ref{degC3})--(\ref{degC5}) and single-polarization 
approximation discussed in Eqs. (\ref{degB12}), (\ref{degB13}) and (\ref{degB19}).
Equation (\ref{degC5a}) suggests that the interference of the intensities of a single-polarization
can be approximated with the interference of the intensity of a single mode of the field. The effect of the polarizations is a progressive reduction of the degree of second-order coherence. This reduction preserves the super-Poissonian character of the quantum state so that the Poissonian limit (typical of the coherent state) is never reached.
\renewcommand{\theequation}{5.\arabic{equation}}
\setcounter{equation}{0}
\section{Stimulated versus spontaneous emission}
\label{sec5}
The stimulated emission of relic gravitons does not reduce the 
degrees of coherence below the Poissonian limit provided the 
average multiplicity of the initial state does not dominate against 
the average multiplicity of the produced gravitons. 
This conclusion partly follows from Eqs. (\ref{degA29}), (\ref{degB18}) and 
(\ref{degB19}) where the average multiplicity of gravitons at $\tau_{i}$ 
has been already considered. When the average multiplicity of the initial state does not vanish 
the previous results describe the interference of the intensities 
of the relic gravitons produced by stimulated emission. Conversely 
in the limit $\overline{n}_{k}(\tau_{i}) \to 0$ the same 
expressions hold in the case of spontaneous emission where 
$\hat{b}_{k}(\tau_{i})$ annihilates the vacuum at $\tau_{i}$.
While the previous analyses show that the super-Poissonian behaviour
 is not altered by more general parametrizations of the initial state, it is also
 true that the parametrization of the initial state suggested above is not the 
 most general one. This complementary aspect of the present analysis will now be 
 clarified by considering the initial conditions provided by a coherent state in the continuous mode representation. Since a given multiparticle density matrix can be projected on the coherent 
state basis via the so-called Klauder-Sudarshan $P$-representation \cite{QO4}
this analysis seems appropriate and sufficiently conclusive, at least for the present 
purposes.

\subsection{Squeezed coherent states}
For a continuum of modes the Glauber displacement operator is defined as \cite{kibble}
\begin{equation}
{\mathcal D}(\alpha) = e^{d(\alpha)}, \qquad d(\alpha) = \sum_{\lambda} \int d^{3} k \biggl[ \alpha_{\vec{k}\,\lambda} \hat{a}^{\dagger}_{\vec{k},\lambda} - \alpha^{*}_{\vec{k}\,\lambda} \hat{a}_{\vec{k},\lambda}\biggr],
\label{coh1}
\end{equation}
with the same notations already employed for the squeezing and rotation operators ${\mathcal S}(z)$ and ${\mathcal R}(\delta)$ of Eqs. (\ref{degA9a}) and (\ref{BB2}). The squeezed coherent states of relic gravitons can be 
introduced in two complementary perspectives mirroring their quantum optical analogs
originally discussed by Caves \cite{caves} (sometimes also referred to as ideal squeezed state) 
an by Yuen \cite{yuen} (the so-called two-photon coherent states). 
In the Caves representation the initial state is rotated,  squeezed and finally dispalced\footnote{We recall that  that, by definition, $| \{z\, \delta \}\rangle = {\mathcal S}(z) \,{\mathcal R}(\delta) |\mathrm{vac}\rangle$ and $| \{ \beta \}\rangle = {\mathcal D}(\beta) |\mathrm{vac}\rangle$. } i.e. $| \{\alpha\, z\, \delta \}\rangle = {\mathcal D}(\alpha) | \{z\, \delta \}\rangle$. In the Yuen representation \cite{yuen} the squeezed-coherent states of relic gravitons are instead defined as $|\{  z\, \delta\, \beta \}\rangle = {\mathcal S}(z) {\mathcal R}(\delta) |\{ \beta\}\rangle$. 
According to the strategy of Ref. \cite{yuen} (appropriately extended to the continuous-mode description 
of Eq. (\ref{coh1})) the creation operators transform as:
\begin{eqnarray}
&& {\mathcal D}^{\dagger}(\beta)\,{\mathcal R}^{\dagger}(\delta) \, {\mathcal S}^{\dagger}(z) \, \hat{a}_{\vec{q}\,\,\lambda}{\mathcal S}(z) \, {\mathcal R}(\delta) 
{\mathcal D}(\beta) = 
\nonumber\\
&& \biggl[e^{- i\, \delta_{q\,\lambda}} \cosh{r_{q\,\lambda}} \beta_{\vec{q}\,\,\lambda}
-  e^{i (\theta_{q\,\lambda} + \delta_{q\,\lambda})} \sinh{r_{q\,\,\lambda}} \beta^{*}_{-\vec{q}\,\lambda}\biggr]
\nonumber\\
&&+ e^{- i\, \delta_{q\,\lambda}} \cosh{r_{q\,\lambda}} \hat{a}_{\vec{q}\,\lambda} -  e^{i (\theta_{q\,\lambda} + \delta_{q\,\lambda})} \sinh{r_{q\,\lambda}} \hat{a}^{\dagger}_{-\vec{q}\,\lambda}.
\label{BB8}
\end{eqnarray}
If the action of the displacement operator precedes the squeezing and the rotation, as suggested in Ref. \cite{caves}, the creation operators transform instead as:
\begin{eqnarray}
&& {\mathcal R}^{\dagger}(\delta) \, {\mathcal S}^{\dagger}(z) \, {\mathcal D}^{\dagger}(\alpha)\, \hat{a}_{\vec{q}\,\,\lambda} {\mathcal D}(\alpha) {\mathcal S}(z) \, {\mathcal R}(\delta) = \alpha_{\vec{q},\lambda}
\nonumber\\
&&+ e^{- i\, \delta_{q\,\,\lambda}} \cosh{r_{q\,\,\lambda}} \hat{a}_{\vec{q}\,\,\lambda} -  e^{i (\theta_{q\,\,\lambda} + \delta_{q\,\,\lambda}}) \sinh{r_{q\,\,\lambda}} \hat{a}^{\dagger}_{-\vec{q}\,\,\lambda}.
\label{BB9}
\end{eqnarray}
Comparing the two expressions of Eqs. (\ref{BB8}) and (\ref{BB9}) we conclude they are not equivalent in general but coincide, in practice, provided $\alpha_{\vec{q},\lambda}$ equals the expression inside the square bracket appearing in Eq. 
(\ref{BB8}), i.e. 
\begin{equation}
\alpha_{\vec{q},\lambda} = e^{- i\, \delta_{q,\lambda}} \cosh{r_{q,\lambda}} \beta_{\vec{q},\lambda} -  e^{i (\theta_{q,\lambda} 
+ \delta_{q,\lambda}}) \sinh{r_{q,\lambda}} \beta^{*}_{-\vec{q},\lambda}.
\label{BB10}
\end{equation}
 Even if Eqs. (\ref{BB10}), (\ref{BB8}) and (\ref{BB9}) are general, for the present purposes it will be sufficient to consider a single polarization   and then contrast the obtained results with the findings of the previous sections.

\subsection{Degrees of coherence}
The first-order and second-order Glauber correlators for a squeezed-coherent state are
\begin{eqnarray}
{\mathcal S}^{(1)}(x_{1},\, x_{2}) &=&  \langle \{ \delta\,z \, \alpha\}| \hat{\mu}^{(-)}(x_{1}) \, \hat{\mu}^{(+)}(x_{2}) | \{ \alpha\, z\, \delta\}\rangle,
\label{INS1}\\
{\mathcal S}^{(2)}(x_{1},\, x_{2}) &=&  \langle \{ \delta\,z \, \alpha\}| \hat{\mu}^{(-)}(x_{1}) \, \hat{\mu}^{(-)}(x_{2}) \hat{\mu}^{(+)}(x_{2}) \, \hat{\mu}^{(+)}(x_{1}) | \{ \alpha\, z\, \delta\}\rangle.
\label{INS2}
\end{eqnarray}
The action 
of ${\mathcal D}(\alpha)$ over $\hat{\mu}^{(-)}(x)$ and $\hat{\mu}^{(+)}(x)$ displaces 
the field operators by their classical value \cite{kibble}: 
\begin{eqnarray}
{\mathcal D}^{\dagger}(\alpha) \hat{\mu}^{(-)}(x) {\mathcal D}(\alpha) &=& \mu_{c}^{*}(x) + \hat{\mu}^{(-)}(x), 
\nonumber\\
{\mathcal D}^{\dagger}(\alpha) \hat{\mu}^{(+)}(x) {\mathcal D}(\alpha) &=& \mu_{c}(x) + \hat{\mu}^{(+)}(x),
\label{INS3}\\
\mu_{c}(x) &=& \frac{1}{(2\pi)^{3/2}} \int \frac{d^{3} k}{\sqrt{ 2 k}} \,\alpha_{\vec{k}} \, e^{- i \vec{k}\cdot\vec{x}},
\nonumber
\end{eqnarray}
where we stress that $\mu_{c}(x)$ is not an operator but a classical field.  
Inserting Eq. (\ref{INS3}) into Eq. (\ref{INS1}) the first-order Glauber correlator is:
\begin{eqnarray}
{\mathcal S}^{(1)}(x_{1}, x_{2}) &=& \mu_{c}^{*}(x_{1}) \mu_{c}(x_{2}) + \langle \{ \delta\,z \}| \hat{\mu}^{(-)}(x_{1}) \, \hat{\mu}^{(+)}(x_{2}) | \{ z\, \delta\}\rangle
\nonumber\\
&=& \mu_{c}^{*}(x_{1}) \mu_{c}(x_{2}) + W_{0}(r, \tau_{1}, \tau_{2}),
\nonumber\\
W_{0}(r, \tau_{1}, \tau_{2}) &=&  \frac{1}{4 \pi^2} \int k \, d k \, j_{0}(k r) v_{k}^{*}(\tau_{1}) \, v_{k}(\tau_{2}). 
\label{INS4}
\end{eqnarray}
The first term in Eq. (\ref{INS4}), analog to the condensate arising in the theory of superfluidity \cite{valatin1,valatin2}, depends on $x_{1}$ {\em and} $x_{2}$. Conversely the second term $W_{0}(r,\tau_{1},\tau_{2})$ is the quantum contribution which is a function of the distance. The explicit form of the HBT correlations is given by
\begin{eqnarray}
{\mathcal S}^{(2)}(x_{1}, x_{2}) &=& \langle \hat{\mu}^{(-)}(x_{1}) \,  \hat{\mu}^{(-)}(x_{2}) \hat{\mu}^{(+)}(x_{1})  \hat{\mu}^{(+)}(x_{2})\rangle +  |\mu_{c}(x_{1})|^2 |\mu_{c}(x_{2})|^2 
\nonumber\\
&+& |\mu_{c}(x_{1})|^2 \langle \hat{\mu}^{(-)}(x_{2})  \hat{\mu}^{(+)}(x_{2}) \rangle 
+ |\mu_{c}(x_{2})|^2 \langle \hat{\mu}^{(-)}(x_{1})  \hat{\mu}^{(+)}(x_{1}) \rangle 
\nonumber\\
&+&  \mu_{c}^{*}(x_{1}) \mu_{c}^{*}(x_{2}) \langle \hat{\mu}^{(+)}(x_{2})  \hat{\mu}^{(-)}(x_{1}) \rangle 
+ \mu_{c}(x_{1}) \mu_{c}(x_{2})  \langle \hat{\mu}^{(-)}(x_{1})  \hat{\mu}^{(+)}(x_{2}) \rangle 
\nonumber\\
&+& \mu_{c}^{*}(x_{1}) \mu_{c}(x_{2})  \langle \hat{\mu}^{(-)}(x_{2})  \hat{\mu}^{(+)}(x_{1}) \rangle 
+ \mu_{c}(x_{1}) \mu_{c}^{*}(x_{2})  \langle \hat{\mu}^{(-)}(x_{1})  \hat{\mu}^{(+)}(x_{2}) \rangle. 
\label{INS5}
\end{eqnarray}
Besides the contribution of the condensate and of the quantum fluctuations (first line of Eq. (\ref{INS5})), we can identify a mixed contribution (second line of Eq. (\ref{INS5})) and two interference terms (third and fourth lines of Eq. (\ref{INS5})) that depend on the mutual phases characterizing the displacement, squeezing and rotation operators. The degree of second-order coherence can then be expressed in the following manner
\begin{eqnarray}
g^{(2)}(x_{1}, x_{2}) -1 &=& \frac{W_{1}(r, \tau_{1}, \tau_{2})}{[|\mu_{c}(x_{1})|^2 +W_{0}(\tau_{1}) ]\,\,[|\mu_{c}(x_{2})|^2 + W_{0}(\tau_{2})]}
\nonumber\\
&+&\frac{ \mu_{c}^{*}(x_{1}) \mu_{c}(x_{2})  W_{2}(r, \tau_{1}, \tau_{2}) 
+ \mu_{c}(x_{1}) \mu_{c}^{*}(x_{2}) W_{2}^{*}(r, \tau_{1}, \tau_{2}) }{[|\mu_{c}(x_{1})|^2 +W_{0}(\tau_{1}) ]\,\,[|\mu_{c}(x_{2})|^2 + W_{0}(\tau_{2})]}
\nonumber\\
&-&  \frac{\mu_{c}^{*}(x_{1}) \mu_{c}^{*}(x_{2}) W_{3}(r, \tau_{1}, \tau_{2})  
+ \mu_{c}(x_{1}) \mu_{c}(x_{2}) W^{*}_{3}(r, \tau_{1}, \tau_{2}) }{[|\mu_{c}(x_{1})|^2 +W_{0}(\tau_{1}) ]\,\,[|\mu_{c}(x_{2})|^2 + W_{0}(\tau_{2})]},
\label{INS7}
\end{eqnarray}
where, from Eq. (\ref{INS4}), we defined  $W_{0}(\tau) = W_{0}(0, \tau, \tau)$ and also
\begin{eqnarray}
W_{1}(r, \tau_{1}, \tau_{2}) &=& \langle \hat{\mu}^{(-)}(x_{1}) \,  \hat{\mu}^{(-)}(x_{2}) \hat{\mu}^{(+)}(x_{1})  \hat{\mu}^{(+)}(x_{2})\rangle - \langle \hat{\mu}^{(-)}(x_{1})  \hat{\mu}^{(+)}(x_{1})\rangle \langle \hat{\mu}^{(-)}(x_{2})  \hat{\mu}^{(+)}(x_{2})\rangle
\nonumber\\
&=& \frac{1}{16 \pi^4} \int k \, d k \, j_{0}(k r) \int p \, d p\, j_{0}(p r)
\biggl[ v_{k}^{*}(\tau_{1}) v_{p}^{*}(\tau_{2}) v_{k}(\tau_{2}) v_{p}(\tau_{1}) 
\nonumber\\
&+& v_{k}^{*}(\tau_{1}) u_{k}^{*}(\tau_{2}) u_{p}(\tau_{2}) v_{p}(\tau_{1}) \biggr],
\label{INS8}\\
W_{2}(r, \tau_{1}, \tau_{2}) &=& \langle \hat{\mu}^{(-)}(x_{2})  \hat{\mu}^{(+)}(x_{1}) \rangle 
= \frac{1}{4\pi^2} \int k \, d k  \, j_{0}(k r) \, v_{k}^{*}(\tau_{2}) v_{k}(\tau_{1}),
\label{INS9}\\
W_{3}(r, \tau_{1}, \tau_{2}) &=& - \langle \hat{\mu}^{(+)}(x_{2})  \hat{\mu}^{(+)}(x_{1}) \rangle 
= \frac{1}{4\pi^2} \int k \, d k  \, j_{0}(k r) \, u_{k}(\tau_{2}) v_{k}(\tau_{1}).
\label{INS10}
\end{eqnarray}
According to Eq. (\ref{INS7}) and thanks to the explicit form of Eqs. (\ref{INS8})--(\ref{INS10}) 
the sign of $g^{(2)}(x_{1}, x_{2}) -1$ may become negative and 
this possibility is well known from quantum optical studies where 
a squeezed state with a strong coherent component may have a sub-Poissonian 
statistics \cite{QO1,loudon,caves,yuen} provided the average 
multiplicity of the coherent component dominates against the 
the squeezing contribution, as already anticipated at the beginning of this section. 
A similar conclusion will be reached hereunder and to investigate the analog phenomenon in the present case we  
consider the  regime   $x_{1} \to x_{2}$ where Eq. (\ref{INS7}) becomes:
\begin{eqnarray}
g^{(2)}(x) - 1 &=& \frac{W_{0}^2(\tau) + |W_{3}(\tau)|^2 }{[ |\mu_{c}(x)|^2 + W_{0}(\tau)]^2} 
\nonumber\\
&+& 2 \frac{W_{0}(\tau)}{[ |\mu_{c}(x)|^2 + W_{0}(\tau)]^2} - \frac{\mu_{c}^{*\,2}(x) W_{3}(\tau) + \mu_{c}^{2}(x) W^{*}_{3}(\tau) }{[ |\mu_{c}(x)|^2 + W_{0}(\tau)]^2}.
\label{INS11}
\end{eqnarray}
The explicit expressions of $W_{0}(\tau)$, $W_{1}(\tau)$, $W_{2}(\tau)$ and $W_{3}(\tau)$
appearing in Eq. (\ref{INS11}) is:
\begin{eqnarray}
W_{0}(\tau) &=& W_{2}(\tau) = \frac{1}{4 \pi^2}\int k \, d k \, |v_{k}(\tau)|^2,
\label{IN12}\\
W_{1}(\tau) &=& W_{0}^2(\tau) + |W_{3}(\tau)|^2, 
\label{IN13}\\
W_{3}(\tau) &=& \frac{1}{4\pi^2} \int k \, d k \, u_{k}(\tau) \, v_{k}(\tau).
\label{INS15}
\end{eqnarray}
While the term appearing in the first line at the right hand side of Eq. (\ref{INS11}) 
is always positive semidefinite, the remaining two terms (in the second line of the same equation)
do not have a definite sign. We can therefore conclude that  $g^{(2)}(x) > 1$ as long as 
the average multiplicity of the produced gravitons exceeds the coherent component of the initial state. 
In particular when $\mu_{c}(x)\to 0$ the case treated in the previous section is recovered and $g^{(2)}(x) \to 3$.  

\subsection{Wavelengths inside and beyond the effective horizon}
If the spectrum of the initial fluctuations is characterized by a given wavelength 
(for instance a thermal wavelength) at $\tau_{i}$, the present value 
of this characteristic scale will be much larger than the Hubble radius 
at the present time unless the total number of efolds $N_{t}$ is very close\footnote{The value of $N_{crit}$ also depends on the post-inflationary history. Conservative estimates suggest
$N_{crit} = 63 \pm15$ \cite{mg,liddle}. In the case of a standard  post-inflationary history $N_{crit} = 63.6 + (1/4) \ln{\epsilon}$. According to some, for the consistency of the inflationary scenarios we must anyway demand that the total number of efolds exceeds $N_{crit}$.}  to the critical number of efolds $N_{crit} = {\mathcal O}(66)$ \cite{mg,liddle}. Even assuming (or tuning) 
$N_{t} \sim N_{crit}$ it seems difficult to conceive an initial state 
that could make the statistics sub-Poissonian. For this purpose we can investigate more carefully the sign 
of $g^{(2)}(x) -1$ when the coherent component dominates over the average 
multiplicity of the produced gravitons. It is useful to introduce the quantities 
$\epsilon_{0}(x) = W_{0}(\tau)/|\mu_{c}(x)|^2 < 1$ and  $\epsilon_{3}(x) = W_{3}(\tau)/|\mu_{c}(x)|^2  < 1$ that are both smaller than one when the coherent component exceeds the squeezed contribution; 
Eq. (\ref{INS11}) can then be written as:
\begin{eqnarray}
g^{(2)}(x) - 1 &=& \frac{\epsilon_{0}^2(\tau) + | \epsilon_{3}(x)|^2}{[1 + \epsilon_{0}(x)]^2} 
\nonumber\\
&+& 2 \frac{\epsilon_{0}(x)}{[ |\mu_{c}(x)|^2 + W_{0}(\tau)]^2} - \frac{e^{ - 2 i \varphi(x)} \epsilon_{3}(x) + e^{ 2 i \varphi(x)}\epsilon_{3}^{*}(x) }{[1 + \epsilon_{0}(\tau)]^2}, 
\label{INS17}
\end{eqnarray}
where $\mu_{c}(x) = e^{i \varphi(x)} \, |\mu_{c}(x)|$ has been separated in its modulus 
and phase. The first term at the right hand side of Eq. (\ref{INS17}) contains contributions ${\mathcal O}(\epsilon_{0}^2)$ and ${\mathcal O}(\epsilon_{3}^2)$ that are subleading in comparison with the remaining two terms (of order $\epsilon_{0}$ and $\epsilon_{3}$ respectively). The sign of $[g^{(2)}(x) - 1]$ will then be determined by the balance of the dominant contributions at the right hand side. If we now recall Eq. (\ref{degA9}) and assume,  for simplicity, that $\varphi$ is constant we can rephrase the dominant contributions of Eq. (\ref{INS17}) as:
\begin{equation}
g^{(2)}(x) - 1 \simeq \frac{1}{2 \pi^2 |\mu_{c}(x)|^2} \int k \, d k\,\biggl[ |v_{k}(\tau)|^2 - \cos{ 2 \zeta_{k}} |u_{k}(\tau)| \, |v_{k}(\tau)| \biggr],
\label{INS18}
\end{equation}
where $ \zeta_{k} = (\varphi - \theta_{k}/2)$. This result implies that $g^{(2)}(x) - 1 < 0$ when 
$\cos{ 2 \zeta_{k}} > |v_{k}|/|u_{k}|$ and provided the coherent component 
exceeds the squeezing contribution. In this limit, however, the large-scale 
fluctuations will simply correspond to the coherent contribution. Thus the coherent component can only exceed the squeezing contribution provided the total number of efolds  is tuned around its critical value.  If we now recall  Eq. (\ref{degA9}) 
the condition $(|v_{k}(\tau)|^2 - \cos{ 2 \zeta_{k}} |u_{k}(\tau)| \, |v_{k}(\tau)|)<0$ becomes
\begin{equation}
\sinh{r_{k}}[ \sinh{r_{k}} - \cosh{r_{k}}] \cos^2{\zeta_{k}} + \sinh{r_{k}}[ \sinh{r_{k}} + \cosh{r_{k}}] \sin^2{\zeta_{k}}
<0.
\label{INS19}
\end{equation}
After simple algebra, Eq. (\ref{INS19}) can also be expressed as: 
\begin{equation}
(e^{ 2 r_{k}} -1) \sin^2{\zeta_{k}} - ( 1 - e^{- 2 r_{k}}) \cos^2{\zeta_{k}} <0,
\label{INS20}
\end{equation}
showing that if $\zeta_{k} = 0$ the inequality is always verified even in the limit $r_{k} \gg 1$.
The results of the Caves approach \cite{caves} discussed so far 
can be translated into the Yuen description \cite{yuen} by making use of Eq. (\ref{BB10})
which can also be written, in the single-polarization approximation, 
\begin{equation}
|\alpha_{k}|^2 = |\beta_{k}|^2 [ \cosh{2 r_{k}} - \sinh{2 r_{k}} \cos{ 2 \gamma_{k}}] = 
|\beta_{k}|^2 \biggl[ e^{ - 2 r_{k}} \cos^2{\gamma_{k}} + e^{ 2 r_{k}} \sin^2{\gamma_{k}} \biggr].
\label{INS21}
\end{equation}
where $\gamma_{k} = [(\delta_{k} +\theta_{k}/2) - \chi_{k}]$.
From Eq. (\ref{BB10}) the relation between $\zeta_{k}$ and $\gamma_{k}$ is given by:
\begin{equation}
\cos^2{\zeta_{k}} = \frac{e^{- 2 r_{k}} \, \cos^2{\gamma_{k}}}{e^{ - 2 r_{k}} \cos^2{\gamma_{k}} + e^{ 2 r_{k}} \sin^2{\gamma_{k}}}, \qquad \sin^2{\zeta_{k}} = \frac{e^{ 2 r_{k}} \, \sin^2{\gamma_{k}}}{e^{ - 2 r_{k}} \cos^2{\gamma_{k}} + e^{ 2 r_{k}} \sin^2{\gamma_{k}}},
\label{INS22}
\end{equation}
which also demands  $e^{4 r_{k}} ( 1 - e^{- 2 r_{k}}) \sin^2\gamma_{k} < e^{- 2 r_{k}}( 1 - e^{- 2 r_{k}}) \cos^2{\gamma_{k}}$. 
In analogy with Eq. (\ref{INS20}),  if $\gamma_{k} \to 0$ the previous inequality 
is always verified even in the limit $r_{k} \gg 1$.  But unfortunately the limit $\gamma_{k} \to 0$ can only be realized in the exact de Sitter case (see Eq. (\ref{degA20}))  or if we consistently tune
 $\chi_{k} \to (\delta_{k} +\theta_{k}/2)$  for all modes that exceed the Hubble radius. Conversely, without 
 fine-tuning, Eq. (\ref{degA24a}) implies $\gamma_{k} \simeq - \pi\epsilon/2$ in the quasi-de Sitter case  and Eq. (\ref{INS22}) demands
\begin{equation}
e^{4 r_{k}} ( 1 - e^{- 2 r_{k}}) \frac{\pi^2 \epsilon^2}{2} - e^{- 2 r_{k}}( 1 - e^{- 2 r_{k}}) \biggl(1 - \frac{\pi^2\epsilon^2}{4}\biggr) < 0.
\label{INS22b}
\end{equation}
In the second term we can always neglect the $\epsilon^2$ correction which is small with respect 
to $1$; from the remaining terms we have 
$1 < r_{k} < (1/6) \ln{[4/(\pi^2\epsilon^2) -1/2]}$ i.e. $r_{k}< 2.15$ for $\epsilon = 0.001$. 
This condition is therefore unphysical since the average multiplicity of the gravitons 
produced during inflation would be negligibly small and this would imply that 
the tensor modes are not amplified in comparison with the scalar modes\footnote{It should be clear 
that the value $\epsilon = 0.001$ has not been randomly chosen. In the concordance paradigm 
the slow-roll parameter epsilon is related to the tensor-to-scalar 
ratio $r_{T}$ (not to be confused with $r_{k}$) as $r_{T} = 16 \epsilon$. Since 
$r_{T}$ must be smaller than $0.07$ (or even $0.05$) \cite{BICPL} 
we also have that epsilon must be conservatively of order $10^{-3}$. The addition of gauge fields 
in the game may increase $r_{T}$ but may also affect the scalar mode so that, in this case, it is possible to show 
that $r_{T}$ must be bounded from below (i.e. $10^{-3} < r_{T} < 1$ \cite{mg1}).}.

The result of Eq. (\ref{INS22b}) is, in some sense, pleonastic since the relic gravitons  potentially observable today (for instance in the audio band) are all inside the effective horizon (i.e.  $k\tau\gg 1$) and, in this limit, $\gamma_{k} \neq 0$ for independent reasons.  Again the 
statistics can never become sub-Poissonian unless the average multiplicity of gravitons produced during inflation is negligible\footnote{The present results are at odds with the claim of Ref. \cite{KS} (obtained in the single-mode approximation) where 
the authors suggest that the analog of $\gamma_{k}$ is generically vanishing. This is only true 
provide the phase of the coherent state is tuned in a way that the squeezed contribution is exactly 
cancelled.}. Inside the Hubble radius  the phases are determined by the pair of conditions 
\begin{eqnarray}
e^{- i \delta_{k}} &=& \frac{c_{+}(k)}{|c_{+}(k)|} e^{- i k\tau} = e^{- i (k \tau + \nu_{+}) - i\pi/2} \frac{[z (i - q^{-1}) + (1 + i) z^{-1}]}{\sqrt{z^2 + (z/q)^2 + 2/z^2 + 2 - 2/q}} 
\nonumber\\
&\to& e^{- i (k \tau + \nu_{+}) + i \pi/2}\frac{ 1 - i q}{\sqrt{q^2 + 1}},
\label{IN16}\\
e^{i (\delta_{k}+ \theta_{k})} &=& - \frac{c_{-}^{*}(k)}{|c_{-}(k)|} e^{- i k\tau} =  e^{- i (k \tau - \nu_{-}) - i \pi/2} \frac{[z (-i + q^{-1} ) - (1 - i)z^{-1}]}{\sqrt{z^2 + (z/q)^2 + 2/z^2 - 2 - 2/q}} 
\nonumber\\
&\to& e^{- i (k \tau -  \nu_{-}) -i \pi/2}\frac{1 - i q}{\sqrt{q^2 + 1}},
\label{IN17}
\end{eqnarray}
where $z= a_{re}/a_{ex}$, $q= k/{\mathcal H}_{re}$ and $\nu_{\pm}(k) = k(\tau_{ex} \mp \tau_{re})$. In  Eqs. (\ref{IN16}) and (\ref{IN17}) the corresponding expressions have been simplified in the limit $ z \gg 1$ and $k/{\mathcal H}_{re} < z^2$ so that the squeezing phases are given by 
\begin{eqnarray}
\delta_{k} &=& k \tau + \nu_{+}(k) - \frac{\pi}{2} + \arctan{q} 
\label{IN18}\\
\theta_{k} + \delta_{k} &=& - k\tau + \nu_{-}(k) - \frac{\pi}{2} -\arctan{q} 
\label{IN19}
\end{eqnarray}
But this means $\delta_{k} + 2 \theta_{k} \simeq [\nu_{+} + \nu_{-}]/2  = k \tau_{ex}= {\mathcal O}(1)$ implying that, for squeezed coherent states, the statistics of the relic gravitons is never sub-Poissonian.  Following the ideas conveyed in this section, different initial 
states such as squeezed-number states or even mixed states (e.g. squeezed 
thermal states) can be analyzed with similar qualitative results. In some cases 
the degree of second-order coherence can be reduced by the presence of an appropriate initial state.
While we leave the explicit analysis of this interesting point to a more topical discussion, 
we can safely conclude that the properties of the initial states may very well interfere with the squeezing 
contribution but do not affect the super-Poissonian character of the degree of second-order 
coherence and of the HBT correlations especially inside the effective horizon\footnote{It has been recently shown 
that the degree of second-order coherence of relic gravitons in a squeezed number state can be smaller than $3$ 
and it goes to $1.5$ when the average multiplicity of the created gravitons is much larger than the average multiplicity 
of the initial state \cite{mg2}. We can therefore say that the degree of second-order coherence is generically between 
$1.5$ and $3$.}. 

Before concluding this section it is useful to recapitulate the overall perspective of the 
present analysis that has been conducted by considering the limit of the relevant 
Glauber correlators when the wavelengths are either 
larger than the effective horizon (sometimes also dubbed Hubble radius in the previous 
sections) or shorter than the effective horizon. The frequency range where these 
two limits are verified depends on the model under consideration. The simplest 
framework where a concrete estimate is possible, as already remarked in the last paragraph of the 
introduction, is represented by the concordance paradigm \cite{BICPL} where the only source 
of inhomogeneity is represented by the scalar and tensor modes of the geometry. This 
is also the perspective conveyed in the first applications of HBT interferometry to cosmology \cite{HBT3,HBT4}.
The concordance paradigm enhanced by the tensor modes is sometimes dubbed $T\Lambda$CDM where 
$T$ stands for the tensor component, $\Lambda$ represents  the dark energy component and 
CDM reminds of the cold dark matter component. This scenario is characterized 
by $7$ independent parameters and the tensor component is described by the 
tensor-to scalar-ratio $r_{T}$ defined in the introduction and controlling, at once, the amplitude of the spectral energy density and its slope. The spectral energy density in critical units (i.e. $h_{0}^2 \Omega_{gw}$) roughly 
decreases as the inverse frequency square for between few aHz and 100 aHz while it is quasi-flat (i.e. slightly decreasing) between 100 aHz and the GHz. As already mentioned in the introduction and in the previous section, 
the spectral energy density in the quasi-flat branch is optimistically ${\mathcal O}(10^{-16.5})$ since the absolute normalization of the tensor power spectrum solely depends on the tensor to scalar ratio $r_{T} < 0.07$ \cite{BICPL}. The corresponding chirp amplitude is ${\mathcal O}(10^{-29})$ for a comoving frequency of ${\mathcal O}(0.1)$ kHz. Of course 
the signal may get larger when the spectral energy density increases for frequencies larger than the mHz as it happens when the tensor modes of the geometry 
inherit a refractive index \cite{blue1,blue2} or in the presence of stiff phases. In these cases, as already mentioned,  it can happen that
$h_{c} = {\mathcal O}(10^{-25})$ \cite{blue1}, while, for comparison, the chirp amplitude$h_{c}$ corresponding 
to the astronomical signals detected so far by the Ligo/Virgo collaboration is ${\mathcal O}(10^{-21})$ \cite{LIGO1,LIGO2,LIGO3}.

Between few aHz and 100 aHz the low frequency branch of the concordance spectrum is universal and it is caused by the tensor modes of the geometry reentering the effective horizon after matter-radiation equality. 
Between 100 aHz and 100 MHz the spectral energy density depends on the modes reentering the 
effective horizon during the radiation-dominated stage of expansion and the related degrees 
of coherence follow, in this frequency range, from the results of sections \ref{sec3}.3, \ref{sec4}.3 and \ref{sec5}.3. 
In these cases  the relevant wavelengths are all shorter than the Hubble radius. We can therefore 
conclude that the relic gravitons, in the concordance paradigm, are always first-order coherent and their degree 
of second-order coherence is always super-Poissonian. The estimates of the degrees 
of quantum coherence for wavelengths larger than the Hubble radius (see sections \ref{sec3}.2, \ref{sec4}.2 and part of section \ref{sec5}.3) can be instead 
applied in the complementary situation where the wave $k \tau < 1$: this is for instance 
the regime where the initial conditions of the Einstein-Boltzmann hierarchy are 
set prior to matter-radiation equality. The estimates of sections \ref{sec3} and \ref{sec4}
show that the degree of second-order coherence is also super-Poissonian  when 
the relevant wavelengths are larger than the Hubble radius and this conclusion is relevant 
for potential tests of the HBT correlations in CMB physics, as already remarked in the past \cite{HBT4} and at the 
beginning of section \ref{sec4}.3. 

In section \ref{sec2} we started the analysis from the  expressions of the Glauber correlators 
written in the tensor case since these expressions have never been discussed before in their 
full generality. The expressions for the derived HBT correlations and for the degrees 
of quantum coherence discussed in sections \ref{sec3} and \ref{sec4} are also general.
The estimates of the degrees of quantum coherence presented in the previous sections 
hold, in practice, also when the initial state is characterized by an average multiplicity of gravitons, as 
remarked at the beginning of this section. In can however 
happen, as stressed prior to Eq. (\ref{INS17}) that the initial state has its own 
degree of coherence. We confirm, after the analysis, that even assuming (or tuning) 
$N_{t} \sim N_{crit}$ it seems difficult to conceive an initial state 
that could make the statistics sub-Poissonian even in the case of an initial Fock state whose 
statistics is notoriously sub-Poissonian \cite{QO1,QO2,QO3,mg2}. 
All in all the analysis of the Hanbury Brown-Twiss correlations shows, in a conservative 
perspective, that the degree of second-order coherence  is always super-Poissonian in the 
context of the concordance paradigm both for the spontaneous and for the 
stimulated emission of the relic gravitons.

\renewcommand{\theequation}{6.\arabic{equation}}
\setcounter{equation}{0}
\section{Concluding discussion}
\label{sec6}
To assess the classical or quantum origin of the relic gravitons the only sound strategy
is a careful scrutiny of the higher degrees of quantum coherence. The first step along this direction 
is a  proper extension of  the Glauber theory of quantum coherence to the case of the 
tensor modes of the geometry and this has been the  aim of the present investigation. 
The degree of first-order coherence of relic gravitons always tends to $1$ when the corresponding  
wavelengths  are either larger or smaller than the effective horizon. In standard 
Young interferometry the degree of first-order coherence goes to $1$ when the 
interference fringes are maximized while it goes to $0$ in the opposite case when the 
field is incoherent. Classical configurations and quantum states of a given optical field lead to 
comparable degrees of first-order coherence and this conclusion remains 
practically unchanged in the case of relic gravitons. The analysis of the Hanbury Brown-Twiss correlations 
in their canonical form shows instead that the degree of second-order coherence is always larger than $1$. In the quantum optical jargon the relic gravitons are therefore bunched and  their statistics is super-Poissonian.  The results are physically similar if more exclusive approaches are adopted
when, for instance, a single tensor polarization or even a single mode contributes to the total intensity of the field. The obtained conclusions do not change if we consider  stimulated (rather than spontaneous) emission of relic gravitons. While the super-Poissonian nature of the degree of second-order coherence is a necessary condition if we want to infer the quantum origin of the relic gravitons, such a requirement is not sufficient since other states (for instance mixed) may lead to a super-Poissonian degree of second-order coherence. 
Even if, according to some, the quantum origin of the relic gravitons and of large-scale curvature perturbations is indisputable,  the spirit of the present analysis  is more modest and it aims at formulating a set of criteria which could independently rule in (or out) the conventional viewpoint. The obtained results suggest the Hanbury Brown-Twiss interferometry and the scrutiny of the higher degrees of quantum coherence could be a valid (and probably unique) tool in these matters.

{\it Note added in proofs}

While correcting the proof of the paper,  a preprint appeared on the
archive by S. Kanno \cite{K2} reinstating the viewpoint already
expressed in \cite{KS}. Reference \cite{K2} suggests that the requirement
$\gamma_{k} =0$ cannot be obtained from the dynamics but should be
extrinsically imposed as a ``necessary condition'' (see statements in
section 4.6 of Ref. \cite{K2} and discussion therein) with the aim of obtaining a
sub-Poissonian statistics. This  condition, well
known from analog quantum optical studies \cite{caves,yuen} (see also \cite{HBT3}),
should follow from the dynamics that suggests instead the
opposite in the case of relic gravitons. The present findings, obtained from
the full Glauber correlators properly defined in the tensor case, show that
the super-Poissonian statistics is always the natural outcome of a potential
observation able to resolve the HBTcorrelations for the relic gravitons in
different kinematical regions (i.e. when the wavelengths are either larger or
smaller than
the effective horizons).  This is true, in particular, in the audio band
(i.e. between few Hz and $10$ kHz).

\section*{Acknowledgements}

It is a pleasure to thank T. Basaglia, A. Gentil-Beccot, M. Medves and S. Rohr of the CERN Scientific Information 
Service for their kind help.

\newpage
\begin{appendix}
\renewcommand{\theequation}{A.\arabic{equation}}
\setcounter{equation}{0}
\section{Quantum theory of parametric amplification}
\label{APPA}
The action describing the parametric amplification of the relic gravitons 
can be compactly expressed as \cite{fordp1}
\begin{equation}
S^{(t)}=  \frac{1}{8 \ell_{P}^2} \int d^{3} x \, \int d\tau \,\,\biggl\{ a^2 \biggl[ \partial_{\tau} h_{ij} \partial_{\tau} h_{ij} - \partial_{k} h_{ij} \partial^{k} h_{ij} \biggr] - 4 \ell_{P}^2 a^4 \Pi^{ij}_{(t)} h_{ij}\biggl\},
\label{AA1}
\end{equation}
where the tensor component of the anisotropic stress has been included for completeness.
The rescaled tensor amplitude $ \mu_{ij} =  a \,h_{ij}$ and the anisotropic stress can be always expressed in terms 
of the corresponding polarizations\footnote{We remind that $e^{(\alpha)}_{ij}$ denotes the two tensor polarizations; in the sums of Eq. (\ref{AA2}) 
the index $\alpha = \otimes, \oplus$ where $e^{\oplus}_{ij} = (\hat{m}_{i} \hat{m}_{j} - \hat{n}_{i} \hat{n}_{j})$ and 
$e^{\otimes}_{ij} = (\hat{m}_{i} \hat{n}_{j} + \hat{m}_{j} \hat{n}_{i})$ are expressed in terms of the 
mutually orthogonal unit vectors $\hat{m}$ and $\hat{n}$ that are also orthogonal to the comoving three-momentum of the wave.}
\begin{equation}
\mu_{ij}(\vec{x},\tau) = \sqrt{2} \, \ell_{P} \sum_{\alpha} \, \mu_{\alpha} \, e^{(\alpha)}_{ij}, 
\qquad \Pi^{(t)}_{ij}(\vec{x},\tau) = \sqrt{2} \, \ell_{P} \sum_{\alpha} \, \Pi^{(t)}_{\alpha} \, e^{(\alpha)}_{ij},
\label{AA2}
\end{equation}
so that the resulting tensor Hamiltonian derived from Eq. (\ref{AA1}) becomes:
\begin{equation}
H^{(t)} = \frac{1}{2} \sum_{\alpha} \int d^{3} x \biggl[ \pi_{\alpha}^2 + 2 {\mathcal H} \mu_{\alpha} \pi_{\alpha} + \partial_{k} \mu_{\alpha} \partial_{k} \mu_{\alpha} + 
4 \ell_{P}^2 a^3 \mu_{\alpha} \Pi^{(t)}_{\alpha}\biggr],
\label{AA3}
\end{equation}
where $\pi_{\alpha} = \mu_{\alpha}^{\prime} - {\mathcal H} \mu_{\alpha}$ 
(not to be confused with the anisotropic stress) denotes the canonical momentum.
The classical fields can then be promoted to the status of field operators 
and then Fourier transformed:
\begin{equation}
\hat{\mu}_{\alpha}(\vec{x}, \tau) = \frac{1}{(2\pi)^{3/2}} \int d^{3} p \, \hat{\mu}_{\vec{p},\,\alpha}(\tau) 
e^{- i\, \vec{p}\cdot\vec{x}}, \qquad 
\hat{\pi}_{\alpha}(\vec{x}, \tau) = \frac{1}{(2\pi)^{3/2}} \int d^{3} p \, \hat{\pi}_{\vec{p},\,\alpha}(\tau) 
e^{- i\, \vec{p}\cdot\vec{x}}.
\label{AA4}
\end{equation}
In terms of $ \hat{\mu}_{\vec{p}\,\alpha}$ and $\hat{\pi}_{\vec{p}\,\alpha}$ 
the Hamiltonian (\ref{AA3}) becomes:
\begin{eqnarray}
\hat{H}^{(t)} &=& \frac{1}{2} \int d^{3} p \sum_{\alpha} \biggl\{ \hat{\pi}_{-\vec{p}\,\,\alpha}\, \hat{\pi}_{\vec{p}\,\,\alpha} + p^2 \hat{\mu}_{-\vec{p}\,\,\alpha}\, \hat{\mu}_{\vec{p}\,\,\alpha} 
+ {\mathcal H} \biggl[ \hat{\pi}_{-\vec{p}\,\,\alpha}\, \hat{\mu}_{\vec{p}\,\,\alpha} + 
\hat{\mu}_{-\vec{p}\,\,\alpha}\, \hat{\pi}_{\vec{p}\,\,\alpha} \biggr] 
\nonumber\\
&+& 2 \ell_{P}^2 \, a^3 \biggl[ \hat{\Pi}^{(t)}_{-\vec{p}\,\,\alpha}\, \hat{\mu}_{\vec{p}\,\,\alpha} + 
\hat{\mu}_{-\vec{p}\,\,\alpha}\, \hat{\Pi}^{(t)}_{\vec{p}\,\,\alpha}\biggr]\biggr\}.
\label{AA5}
\end{eqnarray}
The creation and annihilation operators obey the 
 commutation relation $[\hat{a}_{\vec{k},\,\alpha}, \, \hat{a}_{\vec{p},\,\beta}^{\dagger}] 
= \delta^{(3)}(\vec{k} - \vec{p})\, \delta_{\alpha\beta}$ and are related to $\hat{\mu}_{\vec{p} \, \alpha}$ and 
$\hat{\pi}_{\vec{p} \,\alpha} $ via the following pair of equations: 
\begin{equation}
\hat{\mu}_{\vec{p} \, \alpha} = \frac{1}{\sqrt{2 p}} \biggl[ \hat{a}_{\vec{p},\,\alpha} + \hat{a}^{\dagger}_{-\vec{p},\,\alpha} \biggr], \qquad \hat{\pi}_{\vec{p} \,\alpha} = - i\,\sqrt{\frac{p}{2}} 
\biggl[ \hat{a}_{\vec{p},\,\alpha} - \hat{a}^{\dagger}_{-\vec{p},\,\alpha} \biggr].
\label{AA6}
\end{equation}
Inserting Eq. (\ref{AA6}) into Eq. (\ref{AA5}) we have that the Hamiltonian of the problem is given by:
\begin{eqnarray}
\hat{H}^{(t)} &=& \frac{1}{2} \int d^{3} p \sum_{\alpha} \biggl\{ p \biggl[ \hat{a}^{\dagger}_{\vec{p},\,\alpha} \hat{a}_{\vec{p},\,\alpha} + \hat{a}_{- \vec{p},\,\alpha} \hat{a}^{\dagger}_{-\vec{p},\,\alpha} \biggr] + i {\mathcal H} \biggl[ \hat{a}^{\dagger}_{-\vec{p},\,\alpha} \hat{a}_{\vec{p},\,\alpha}^{\dagger} 
-  \hat{a}_{\vec{p},\,\alpha} \hat{a}_{-\vec{p},\,\alpha}\biggr]
\nonumber\\
&+& \gamma_{-\vec{p}\, \alpha} \hat{a}_{\vec{p}\,\alpha} + \gamma_{-\vec{p}\, \alpha}^{*} \hat{a}_{\vec{p}\,\alpha}^{\dagger}
+  \gamma_{\vec{p}\, \alpha} \hat{a}_{- \vec{p}\,\alpha} +  \gamma_{\vec{p}\,\alpha}^{*} \hat{a}_{- \vec{p}\,\alpha}^{\dagger} \biggr\},
\label{AA7}
\end{eqnarray}
where $\gamma_{\pm\vec{p}\,\alpha}$ is related to the presence of the anisotropic stress and may 
lead to a coherent component that has been specifically discussed in section \ref{sec5} 
in general terms. 
The Hamiltonian for the scalar modes of the geometry has the same form of Eq. (\ref{AA7}) 
but a different pump field \cite{HBT3,HBT4}. The Hamiltonian of the  problem 
can be phrased as 
\begin{eqnarray}
\hat{H}^{(t)} &=& \frac{1}{2} \int d^{3} p \sum_{\alpha} \biggl\{ p \biggl[ \hat{a}^{\dagger}_{\vec{p},\,\alpha} \hat{a}_{\vec{p},\,\alpha} + \hat{a}_{- \vec{p},\,\alpha} \hat{a}^{\dagger}_{-\vec{p},\,\alpha} \biggr] 
+\lambda \hat{a}^{\dagger}_{-\vec{p},\,\alpha} \hat{a}_{\vec{p},\,\alpha}^{\dagger} 
+ \lambda^{*} \hat{a}_{\vec{p},\,\alpha} \hat{a}_{-\vec{p},\,\alpha}
\nonumber\\
&+& \gamma_{-\vec{p}\, \alpha} \hat{a}_{\vec{p}\,\alpha} + \gamma_{-\vec{p}\, \alpha}^{*} \hat{a}_{\vec{p}\,\alpha}^{\dagger}
+  \gamma_{\vec{p}\, \alpha} \hat{a}_{- \vec{p}\,\alpha} +  \gamma_{\vec{p}\,\alpha}^{*} \hat{a}_{- \vec{p}\,\alpha}^{\dagger} \biggr\},
\label{AA8}
\end{eqnarray}
and $ \lambda = i {\mathcal H}$. Equation (\ref{AA8}) describes the  parametric amplification of relic gravitons and its quantum optical analog has been firstly discussed in \cite{PP1} in the single-mode approximation. Equation
(\ref{AA8}) may also describe an interacting bose gas at zero temperature \cite{fetter,solomon,valatin1,valatin2} and, in this case, the free  Hamiltonian corresponds to the kinetic energy while the interaction terms account for the two-body collisions with small momentum transfer. In a cosmological context the one-mode analog of Eq. (\ref{AA8}) 
is implicit in Refs. \cite{GR3,sq2}; the quantum theory of parametric amplification both in the scalar and in the tensor case has been firstly discussed in connection with the HBT interferometry in Refs. \cite{HBT3,HBT4}.
The evolution equations in the Heisenberg description follow from the Hamiltonian (\ref{AA8}) and are: 
\begin{eqnarray}
\frac{d \hat{a}_{\vec{p}\,\,\alpha}}{d\tau} &=& - i\, p \, \hat{a}_{\vec{p},\,\alpha} -  i \, \lambda \hat{a}_{- \vec{p},\,\alpha}^{\dagger} - i \gamma_{-\vec{p},\,\alpha}^{*},
\nonumber\\
\frac{d \hat{a}_{-\vec{p}\,\,\alpha}^{\dagger}}{d\tau} &=&  i\, p \, \hat{a}_{-\vec{p},\,\alpha}^{\dagger} +  i \, \lambda^{*} \hat{a}_{\vec{p},\,\alpha} + i \gamma_{\vec{p},\,\alpha}.
\label{AA9}
\end{eqnarray}
The Hamiltonian (\ref{AA8}) can be diagonalized via a canonical transformation of the type \cite{fetter}
\begin{eqnarray}
\hat{a}_{\vec{p},\,\alpha}(\tau) &=& u_{p,\,\alpha}(\tau,\tau_{i}) \hat{b}_{\vec{p},\,\alpha}(\tau_{i}) -  
v_{p,\,\alpha}(\tau,\tau_{i}) \hat{b}_{-\vec{p},\,\alpha}(\tau_{i})  + \zeta_{\vec{p},\, \alpha}, 
\label{AA10}\\
\hat{a}_{-\vec{p},\,\alpha}^{\dagger}(\tau) &=& u_{p,\,\alpha}^{*}(\tau,\tau_{i}) \hat{b}_{-\vec{p},\,\alpha}^{\dagger}(\tau_{i})  -  v_{p,\,\alpha}^{*}(\tau,\tau_{i}) \hat{b}_{\vec{p},\,\alpha}(\tau_{i})  
+ \zeta_{-\vec{p},\, \alpha}^{*}(\tau,\tau_{i}).
\label{AA11}
\end{eqnarray}
Inserting Eqs. (\ref{AA10}) and (\ref{AA11}) into Eq. (\ref{AA9}) the 
evolution equations for $u_{p\,\,\alpha}(\tau,\tau_{i})$ and $v_{p\,\,\alpha}(\tau,\tau_{i})$  are:
\begin{eqnarray}
\frac{d u_{p,\,\alpha}}{d\tau} &=& - i p\, u_{p,\,\alpha}  + i \lambda  v_{p,\,\alpha}^{\ast}, 
\label{AA12a}\\
\frac{d v_{p,\,\alpha}}{d\tau} &=& - i p\, v_{p,\,\alpha} + i \lambda u_{p,\,\alpha}^{\ast},
\label{AA12b}\\ 
\frac{d \zeta_{\vec{p}\,\alpha}}{d\tau} &=& - i p\,\zeta_{\vec{p}\,\alpha} - i \lambda \zeta_{-\vec{p}\,\alpha}^{*} - i \gamma_{-\vec{p}\,\alpha}^{*}.
\label{AA12c}
\end{eqnarray}
The transformation of Eqs. (\ref{AA10}) and (\ref{AA11}) 
preserves the commutation relations between the two different sets of operators provided
$|u_{p\,\,\alpha}(\tau,\tau_{i})|^2 - |v_{p\,\,\alpha}(\tau,\tau_{i})|^2 =1$. The latter condition
holds independently for each mode and for each polarization.
In this case the two complex functions $u_{p\,\,\alpha}(\tau,\tau_{i})$
and $v_{p\,\,\alpha}(\tau,\tau_{i})$ must therefore depend, 
for each polarization and for each mode, upon three real functions (i.e. two phases and one amplitude)
and have been parametrized in terms of the a squeezing 
 amplitude $ r_{p,\alpha}$ supplemented by two squeezing phases (i.e. $\theta_{p,\alpha}$ and $\delta_{p,\alpha}$),
as already mentioned in Eqs. (\ref{degA9}) and (\ref{degA8a})--(\ref{degA8b}). 
Inserting therefore Eq. (\ref{degA9}) into Eqs. (\ref{AA12a}) and (\ref{AA12b}) the evolution of the squeezing parameter and of the phases can be written by:
\begin{eqnarray}
r_{p,\alpha}^{\prime} &=& - {\mathcal H} \cos{\theta_{p,\alpha}}, \qquad \delta_{p,\alpha}^{\prime} = p - {\mathcal H} \sin{\theta_{p,\alpha}} \tanh{r_{p,\alpha}}, 
\label{BB8a}\\
\theta_{p,\alpha}^{\prime} &=& - 2 p  + 2 \frac{{\mathcal H} \sin{\theta_{p,\alpha}}}{\tanh{2 r_{p,\alpha}}}.
\label{BB8b}
\end{eqnarray}
While Eqs. (\ref{AA12a})--(\ref{AA12b}) are linear and can be solved in various cases, the nonlinear equations 
(\ref{BB8a}) and (\ref{BB8b}) are fully equivalent but more difficult to approximate. It would be tempting, for instance, to argue that the limiting form 
of Eqs. (\ref{BB8a})--(\ref{BB8b}) gives the limit of a certain solution of Eqs. (\ref{AA12a})--(\ref{AA12b}). This kind of inference can be however dangerous. For instance Eqs. (\ref{degA21a}), (\ref{degA21b}) and (\ref{degA21c}) 
satisfy Eqs. (\ref{BB8a})--(\ref{BB8b}) in the exact de Sitter case (i.e. ${\mathcal H} = -1/\tau$). In the limit 
$|k \tau | \ll 1$ we have, approximately, that $\delta_{k}+ \theta_{k}/2 \simeq 0$ which is indeed the correct 
result. However, by summing up Eqs. (\ref{BB8a}) and (\ref{BB8b}) and by taking the limit of the obtained expression we would be led to conclude that the same result also holds for $| k\tau | \gg 1$ (i.e. inside the effective horizon). This however does not happen as it can be verified from the exact solution  of Eqs. (\ref{degA21a})--(\ref{degA21c}) implying that the wanted combination is given by:
\begin{equation}
\delta_{k}^{\prime} + \frac{\theta_{k}^{\prime}}{2} = \frac{4 k^3 \tau^2}{4 k^2 \tau^2 +1}.
\label{BB8c}
\end{equation}
If we now take the limit $k \tau \ll 1$ we will have from Eq. (\ref{BB8c}) that $\delta_{k} + \theta_{k}/2 \simeq
4 k^3 \tau^3/3 \ll 1$ that coincides with Eq. (\ref{degA220}). In the opposite limit (i.e. $|k \tau| \gg 1$) Eq. (\ref{BB8c}) 
does not imply $\delta_{k} + \theta_{k}/2 \ll 1$ but rather $\delta_{k} + \theta_{k}/2 \simeq k \tau \gg 1$. 
This result indeed agrees with the exact solution (\ref{degA21a})--(\ref{degA21c}) in the limit 
$| k \tau | \gg1$ (see also Eq. (\ref{lim})).

\renewcommand{\theequation}{B.\arabic{equation}}
\setcounter{equation}{0}
\section{Evolution inside the effective horizon}
\label{APPB}
The relic gravitons potentially observable today are all inside the effective horizon
and to estimate their degrees of quantum coherence the full expressions 
of $u_{k}(\tau)$ and $v_{k}(\tau)$ must be evaluated for $\tau > \tau_{re}$ (see e.g. Eq. (\ref{lim}) and discussion thereafter). For this purpose the idea is to estimate the amplification of the 
mode functions and then 
relate the obtained result to the asymptotes of  $u_{k}(\tau)$ and $v_{k}(\tau)$ in the limit $k \tau \gg 1$ and $\tau > \tau_{re}$. As already mentioned in the bulk of the paper, Eqs. (\ref{AA12a}) 
and (\ref{AA12b}) can be decoupled. In particular the combination $ (u_{k} - v_{k}^{*})/\sqrt{2k} = f_{k}$
obeys the standard equations $f_{k}^{\prime\prime} + [k^2 - a^{\prime\prime}/a] f_{k} =0$ whose solution 
is well known  in the different asymptotic regimes (see e.g. \cite{blue1}). A given wavelength exits the effective horizon (also dubbed sometimes Hubble radius \cite{SW1}) at some typical conformal time $\tau_{ex}$ during 
an inflationary stage of expansion and approximately reenters at $\tau_{re}$, when the 
Universe still expands but in a decelerated manner. The two typical times 
$\tau_{ex}$ and $\tau_{re}$ are the turning points of the WKB approximation \cite{GR3}
and are both determined from the condition $k^2 = |a^{\prime\prime}/a|$. 
If  $|a^{\prime\prime}/a| \neq 0$ in the vicinity 
of the turning point, then $k \tau \simeq 1$ and this is what normally happens at $\tau_{ex}$.
However, when $|a^{\prime\prime}/a| \to 0$ in the region where the turning point is located, the situation is different.
Since $|a^{\prime\prime}/a| \to 0$ during radiation, if the given mode reenters 
during the radiation epoch (or anyway in a regime where the pump field vanishes either approximately or exactly) then $k \tau_{re} \ll 1$ (even if, for $\tau > \tau_{re}$, $|k \tau| >1$).
In the WKB approximation we therefore have 
\begin{eqnarray}
f_{k}(\tau) &=& \frac{1}{\sqrt{2 k}} e^{- i k\tau} , \qquad \tau <\tau_{ex}, 
\label{CC1}\\
f_{k}(\tau) &=& \frac{a}{a_{ex}} \biggl\{ f_{k}(\tau_{ex}) + g_{k}(\tau_{ex}) \int _{\tau_{ex}}^{\tau} 
\frac{a_{ex}^2}{a^2(\tau_{1})} d\tau_{1}
\nonumber\\
&-& k^2 \int_{\tau_{ex}}^{\tau} \frac{d\tau_{1}}{a^2(\tau_{1})} \int_{\tau_{ex}}^{\tau_{1}} a_{ex} 
a(\tau_{2}) f_{k}(\tau_{2}) \, d\tau_{2}\biggr\} , \qquad \tau_{ex} < \tau< \tau_{re}.
\label{CC2}
\end{eqnarray}
Note that $g_{k} = f_{k}^{\prime} - {\mathcal H} f_{k}$ and 
\begin{equation}
g_{k}(\tau) = \frac{a_{ex}}{a} g_{k}(\tau_{ex}) - \frac{k^2}{a(\tau)} \int_{\tau_{ex}}^{\tau} 
a(\tau_{1}) f_{k}(\tau_{1}) \, d\tau_{1}.
\end{equation}
When the modes reenter the Hubble radius (i.e. for $\tau> \tau_{re}$)
the mode function is expressible as $f_{k}(\tau) = [ c_{+}(k) e^{- i k\tau} + c_{-}(k) e^{i k\tau}]/\sqrt{2 k}$
where $c_{\pm}(k)$ are fixed by continuity and are given by\footnote{We correct here a sign 
typo in the overall phase of $c_{\pm}(k)$ that is actually opposite of the one derived in the second paper of Ref. \cite{blue1}.}:
\begin{equation}
c_{\pm}(k) = \frac{ e^{- i k (\tau_{ex} \mp \tau_{re})}}{2 i k} \biggl[ \frac{a_{re}}{a_{ex}} ( i k \mp {\mathcal H}_{re}) + 
\frac{a_{ex}}{a_{re}}( ik + {\mathcal H}_{ex})  
\pm a_{re} \, a_{ex} ({\mathcal H}_{re} \mp ik) ({\mathcal H}_{ex} + i k)  {\mathcal J}(\tau_{ex}, \tau_{re})\biggr],
\label{CC4}
\end{equation}
where  ${\mathcal J}(\tau_{ex}, \tau_{re}) = \int_{\tau_{ex}}^{\tau_{re}} d\tau /a^2(\tau)$ and 
 $c_{\pm}(k)$ satisfy $|c_{+}(k)|^2 - |c_{-}|^2 =1$. We can now go back to the original 
quantities, namely $u_{k}(\tau)$ and $v_{k}(\tau)$. In this way Eqs. (\ref{degA25}) and (\ref{degA26}) 
can be easily obtained. Note furthermore that Eq. (\ref{degA27})  follows from Eq. (\ref{CC4}) 
by simply setting $k \tau_{ex} \simeq 1$ and by keeping the dominant terms without 
violating the conditions $|c_{+}(k)|^2 - |c_{-}|^2 =1$.
 \end{appendix}

\end{document}